\documentclass[showkeys,aps,10pt,twocolumn,showpacs,preprintnumbers,amsmath,amssymb,prd,letterpaper,floatfix,nofootinbib,superscriptaddress, ]{revtex4-1}
\usepackage{graphics}
\usepackage{graphicx}
\usepackage{epsfig}
\usepackage[usenames]{xcolor}
\usepackage{longtable}
\PassOptionsToPackage{hyphens}{url}
\usepackage{hyperref}
\usepackage{soul}
\usepackage{breakurl}
\usepackage{latexsym}
\usepackage{amsmath}
\usepackage{amsthm}
\usepackage{amsfonts}
\usepackage{amssymb}
\usepackage{dsfont}
\usepackage{bm}  
 \newcommand{\be}{\begin{equation}}
\newcommand{\ee}{\end{equation}}
\newcommand{\bel}{\begin{align}}
\newcommand{\eel}{\end{align}}
\newcommand{\bea}{\begin{eqnarray}}
\newcommand{\eea}{\end{eqnarray}} 
\usepackage{nicefrac}
\usepackage{comment}  
\usepackage{epsfig}
\usepackage{color}
\usepackage{latexsym}
\usepackage{varioref}
\usepackage{hyperref}

\usepackage{dsfont}
\usepackage{verbatim}
\usepackage{bm}
\usepackage{bbold}

\renewcommand{\ref}[1]{\autoref}

\begin{document}

\title{ \texorpdfstring{Nucleon-$J/\psi$ and nucleon-$\eta_{c}$ scattering in $P_{c}$  pentaquark channels from LQCD}{}}

\author{U. Skerbis} 
\email{ursa.skerbis@ijs.si}
\affiliation{Jozef Stefan Institute, 1000 Ljubljana, Slovenia}
\author{S. Prelovsek,}
\email{sasa.prelovsek@ijs.si}
\affiliation{Department of Physics, University of Ljubljana, 1000 Ljubljana, Slovenia}
\affiliation{Jozef Stefan Institute, 1000 Ljubljana, Slovenia}
\affiliation{Instit\"ut f\"ur Theoretische Physik, Universit\"at Regensburg, D-93040 Regensburg, Germany}

 \date{\today}

\begin{abstract} 
The lattice QCD simulation of $NJ/\psi$ and $N\eta_c$ scattering is performed at $m_\pi\simeq 266~$MeV in channels with all possible $J^P$. This  includes $J^P=3/2^\pm$ and $5/2^\pm$ where LHCb discovered $P_c(4380)$ and $P_c(4450)$ pentaquark states in proton$-J/\psi$ decay.   This is the first lattice simulation  that reaches the energies $4.3-4.5~$GeV where pentaquarks reside. Several decay channels are open in this energy region and we explore the fate of $P_c$ in the one-channel approximation in this  work. Energies of eigenstates are extracted for the nucleon-charmonium system  at zero total momentum for  all quantum numbers, i.e. six lattice irreducible representations. No significant energy shifts are observed. The number of the observed  lattice eigenstates agrees with the number expected  for non-interacting charmonium and nucleon.  Thus, we do not find any strong indication for a resonance or a bound state in these exotic channels within one-channel approximation. This possibly indicates that the coupling of $NJ/\psi $ channel with other two-hadron channels might be responsible for $P_c$ resonances in experiment.   One of the challenges of this study is that  up to six   degenerate $J/\psi(p) N(-p)$ eigenstates  are expected in the non-interacting limit  due non-zero spins of $J/\psi$ and $N$, and we establish all of them in the spectra. 
 \end{abstract}


\maketitle

\section{Introduction}

 Gell-Mann indicated in his $1964$ paper  \cite{Gell_Mann_1964} that along simple baryons ($qqq$)  and mesons ($q\bar{q}$) also states with more complex structure could exist. This could be particles with mesonic quantum numbers   $q\bar{q}q\bar{q} $ (tetraquarks) or states with baryonic quantum numbers  $qqqq\bar{q}$ or $qqqqqq$  (pentaquarks and dibaryons). In recent years a wide variety of those states were observed by several experiments. Until 2015,  almost all of the  experimentally  confirmed exotic hadrons -so called XYZ states-  carried mesonic quantum numbers. The  confirmed exception with baryonic quantum numbers is dibaryon deuteron.   

In $2015$,  two peaks in  proton-$J/\psi$ invariant mass   with  minimal flavor structure of $uudc\bar{c}$ were observed by  LHCb \cite{LHCb_Pc_2015}.  Discovery was later confirmed by the model independent study  in $2016$ by the same collaboration \cite{2016}.  Two resonances were observed: the lighter with the mass   $M_1\simeq 4380~ \text{MeV}$  has broad width  $\Gamma_2\simeq 205  ~\text{MeV}$, while the heavier state with    $M_2\simeq 4450 ~\text{MeV}$ is narrower with   $ \Gamma_2\simeq 40 ~\text{MeV}.$  Resonances were identified as hidden charm pentaquarks $P_{c}$.  LHCb finds  the best fit for spin-parity assignments $(J_1^{P_1},J_2^{P_2})=(\frac{3}{2}^- ,\; \frac{5}{2}^+)$, while acceptable solutions are also found for additional cases with opposite parity, either ($\frac{3}{2}^+ , \;\frac{5}{2}^-$) or
($\frac{5}{2}^+,\; \frac{3}{2}^-$). 
 
Strong interaction allows $P_{c}\simeq uudc\bar{c}$ to decay  to a charmonium ($\bar cc$) and a nucleon  ($uud$), as well as to different combinations of a charmed meson ($q\bar c$) and a charmed charmed baryon ($qqc$). Some of the allowed decay channels are summarized in Table \ref{tab:particles_pentaquark} in appendix \ref{ch:possible_decays}.

There were several attempts to explain the origin and the structure of  two charmed pentaquarks phenomenologically.  They were predicted   in the coupled channel dynamics   in \cite{Wu2011_PhysRevC.84.015202,Wu2010_PhysRevLett.105.232001},  where the  significant component $\bar D^*\Sigma_c$ couples also to $NJ/\psi$ via the vector meson exchange.  They were studied as hadronic molecules of a charmed meson and a charmed baryon, for example in  \cite{1507.05200_He2015, 1607.03223_He2016, 1808.05815_Rathaud2018,1603.04672_Shen2016,1511.04845_Wang2015c,1806.10384_Wang2018b}, or charmonium and nucleon, for example in \cite{1512.01902_Kahana2015}. All listed studies report to find an indication for   $P_{c}$.  A review of hadronic molecules  including $P_{c}$ can be found in \cite{RevModPhys.90.015004}.   These studies are often based on the phenomenological   meson-exchange models \cite{1611.07107_He2016a, Liu2018, 1510.06271_Lue2015,1508.01468_Wang2015d}. Several studies considered compact diquark-diquark-antiquark internal structure and find $P_c$, for example \cite{Maiani2015a,1507.05867_Lebed2015,1507.07652_Anisovich2015,1711.10736,1507.04950_chic1_p_Guo}.     The heavier $P_c$ was shown to be compatible with the  kinematical effects of rescattering from $N\chi_{c1}$ to $NJ/\psi $
 \cite{1507.04950_chic1_p_Guo}.  Other attempts to explain $P_c$  as a coupled channel effect consider rescattering with a charmed meson and a charmed baryon \cite{1603.02376_Shimizu2016,1709.00819_Yamaguchi2017,Xiao2015_1508.00924}.    Enhancements corresponding to $P_c$ were also found as a consequence of the triangle singularity transitions in the  kinematics of $\Lambda_b   \to K^- pJ/\psi$ \cite{1507.04950_chic1_p_Guo,Liu2016_1507.05359}.  
   Further phenomenological studies can be found in the recent reviews \cite{Chen:2016qju,Lebed:2016hpi,RevModPhys.90.015004,Esposito:2016noz,Ali:2017jda}.

 On the other hand,  there is no knowledge on $P_c$ resonances based on the first-principle lattice QCD at present.
  The lattice simulations of systems with flavor $\bar ccuud$ have never reached energies $4.3-4.5~$GeV where the pentaquarks reside. They  mostly considered interactions of a nucleon and a charmonium near threshold. 
  Slightly attractive interaction was found  in the preliminary study  of $NJ/\psi $ and $N\eta_c$ scattering for $m_\pi=293-598 \text{ MeV}$, where no bound state or resonances appear  \cite{Liu:2008rza_0810.5412}. Similar conclusion was previously  in the quenched simulation   \cite{hep-lat_0605009_Yokokawa:2006td}   at $m_{\pi} \simeq 197 \text{ MeV}$.   Interestingly, the dynamical NPLQCD   study of $N\eta_{c}$ scattering in $s$-wave \cite{Netac} rendered a bound state   about $ 20 ~\text{MeV}$ below threshold for very heavy $m_\pi \simeq 800~$MeV. 
 The simulation  \cite{Jpsi_nucleon_HALQCD_method} presents preliminary results for 
$NJ/\psi $ and $N\eta_c$   potentials and phase shifts   in s-wave using HALQCD method in one-channel approximation. These   were extracted up to the energies $0.2~$GeV above threshold at a very heavy  $m_{u/d}$.     Attractive interaction was found in all channels explored, but not attractive enough to form  bound states or resonances.    Similar approach and conclusions were obtained in an earlier study 
\cite{Kawanai:2010ev,Kawanai:2010cq_1007.1515}. 
 The hadroquarkonium picture was considered in \cite{1608.06537_Alberti2016}, where the static $\bar cc$ potential  $V(r)$ was extracted for $m_c\to \infty$ as function of distance $r$  in the presence of the nucleon. The potential is found shift down only by a few MeV due to the presence of the nucleon.   

  Effective field theory for s-wave quarkonium-nucleon  system near threshold is developed in \cite{Castella2018_1803.05412}, where low-energy constants are determined from lattice data   \cite{ hep-lat_0605009_Yokokawa:2006td,Liu:2008rza_0810.5412,Kawanai:2010ev,Kawanai:2010cq_1007.1515}.   This approach also does not feature bound states or near-threshold resonances. 
   
The aim of this lattice  study is to establish whether  $P_c$ resonances could appear in the one-channel approximation for $NJ/\psi$ or $N\eta_c$ scattering.   This is the first lattice study that reaches energies $4.3-4.5~$MeV where $P_c$ resonances are found in experiment. 
We neglect  contributions from other channels and coupled-channel effects. This study could shed light on whether the   experimental $P_c$ resonances   are crucially  related to the coupled-channel effects or not.    The  purpose is to calculate the eigen-energies of the interacting $NJ/\psi $  system on the lattice in the rest frame for all possible $J^P$. This includes previously studied partial wave $\ell=0$ and for the first time also $\ell>0$. The eigen-energies are then  compared with (i)  the energies expected in the limit when $N$ and $J/\psi$ do not interact  and (ii) the energies expected based on the experimental $P_c$ resonances in the one-channel approximation.  Analogous approach is followed for the  $N\eta_c $ system. The results shed light on the fate of $P_c$ in the one-channel approximation.

\begin{figure}[h!]
	\centering
	\includegraphics[width=0.99\linewidth]{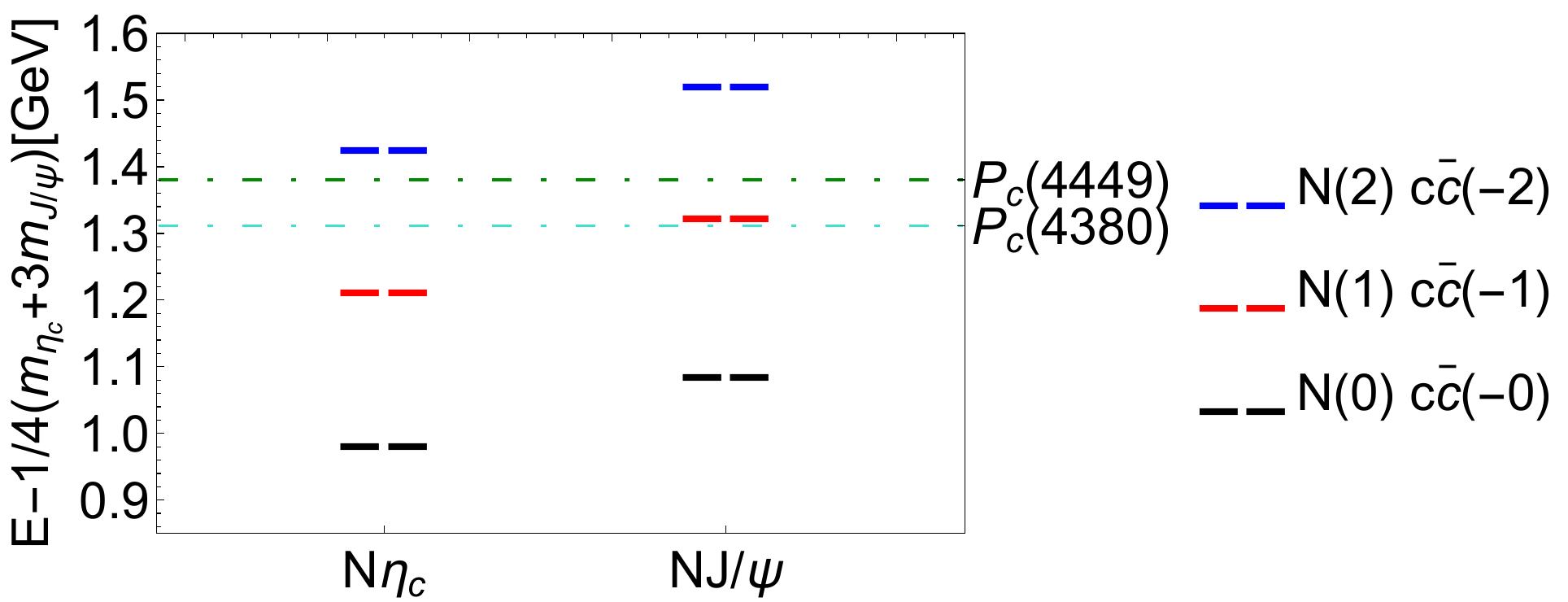}
	\caption{Dashed lines represent the non-interacting energies $E^{n.i.}=E_{N}(p) + E_{\bar cc}(-p) $  (\ref{eqn:non_intracting_energy}) for  the nucleon-charmonium system with total momentum zero on our lattice: black for $p^2=0$, red for $p^2=1$ and blue for $p^2=2$, where $p^2$ is given in units of $(2\pi/{\tt L})^2$.  The experimental masses of  both $P_c$ are also given.  Energies are presented with respect to spin-averaged charmonium mass $(m_{\eta_c}+3 m_{J/\psi})/4$ (\ref{eref}).}
	\label{fig:energy}
\end{figure}

 Let us discuss the expectation for the spectra of $NM$ system ($M=J/\psi$  or  $\eta_c$) in the limit when hadrons do not interact, as this will be an important reference case. The momenta ${\mathbf p}={\mathbf n} \tfrac{2\pi}{\tt L}$ of each hadron are discretized due to the periodic boundary conditions    on our lattice with  ${\tt L}\simeq 2~$fm (\footnote{Symbol ${\tt L}$ denotes the lattice size, while $\ell$ denotes partial wave.}). The total energies of non-interacting (n.i.)  $N(p)M(-p)$ system   are   
 \begin{equation}
\label{eqn:non_intracting_energy}
E^{n.i.} =E_{N}(p)+E_M(-p)\ , \quad \mathbf{p}=\mathbf{n} \tfrac{2\pi}{\tt L} ,\quad \Delta E=E-E^{n.i.}
\end{equation}
with  $\mathbf{n}\in N^3$, while $\mathbf{p}$ with be denoted by $p$ for simplicity from here on. 
The $E_{H=N,M}(p)$  are single-hadron energies measured for various momenta on our lattice (Table 1);  they would  satisfy $E_{H}(p)=(m_H^2+p^2)^{1/2}$  in continuum.      
The non-interacting   energies (\ref{eqn:non_intracting_energy}) for both channels are  presented in Fig.  \ref{fig:energy}, together with the location of $P_c$ resonances from experiment. In order to capture the resonances region, we explore both channels up to   the energy region which captures the states with relative momenta $p^2\leq 2\; (2\pi/{\tt L})^2$ and slightly above. The $NJ/\psi$ scattering is investigated up to  approximately $ E\leq m_{sa}+1.6~$GeV, while  $N\eta_c$  is studied up to $E\leq m_{sa}+1.5~$GeV, where  reference energy is $m_{sa}=(m_{\eta_c}+3 m_{J/\psi})/4$  (\ref{eref}).   

 Furthermore, it is important to understand the effect of the non-zero spins of the scattering particles $N$ and $J/\psi$ in the non-interacting limit.  Even in the continuum infinite-volume theory, several linearly-independent eigenstates have degenerate energy in a given channel $J^P$.   Those are eigenstates with different partial waves $\ell$ and total spins $S$ that render certain $J^P$.  
 For $N\; J/\psi$ channel with $J^P=\frac{1}{2}^+$, for example, there are two linearly independent physical states  
   \bea
\label{eqn:degeneracy_continuum}
J^{P}=\frac{1}{2}^{+} \leftarrow \ell=1, \quad S=1/2  \\
\quad 	\ \quad  \leftarrow  \ell=1, \quad S=3/2 \nonumber
\eea
that have the same energy (\ref{eqn:non_intracting_energy}) in the non-interacting limit and exist for relative momenta $p\!>\!0$ ($p\!=\!0$ does not render partial waves  $\ell>0$). Our aim is to find all  linearly-independent eigenstates and  we will indeed confirm this pair of nearly degenerate eigenstates \footnote{They  appear in irrep $G_1^+$ that contains   $J^P=\frac{1}{2}^+$. There are two levels with $p\!=\!1$  in $G_1^+$ of Fig. \ref{fig:vm_nuk_energy_spectrum}.}.  The effect of finite lattice implies that several  channels $J^P$ contribute to a given lattice irreducible representation (irrep), which further enhances the number of degenerate eigenstates $N(p)M(-p)$ at given $p$. Up to six degenerate eigenstates are expected at nonzero relative momenta for the $N\;J/\psi$ channel and up to two are expected for $N\; \eta_c$. One of the challenges is to extract all linearly independent eigenstates and we indeed establish all of them in our simulation, as will be evidenced from high degeneracies in the Figs.  \ref{fig:vm_nuk_energy_spectrum} and \ref{fig:pm_nuk_energy_spectrum}.

 The existence of a pentaquark resonance $P_c$ would modify the eigen-energies of the nucleon-charmonium system with respect to the non-interacting case discussed above. We address the fate of $P_c$ in one-channel approximation, i.e. when a given pentaquark  resonance couples only to   $N\;J/\psi$ or only to $N\eta_c$, while it is decoupled from other two-hadron channels. If such a narrow resonance $P_c(4450)$ exist with a given spin-parity $J^P$,   the L\"uscher formalism predicts one additional eigenstate  with respect to the non-interacting case in the irreducible representations  that contains this $J^P$. This extra   state  is expected to have an energy approximately    $M_{P_c}\pm \Gamma_{P_c}$, while  nearby states can get shifted in energy. Analogous expectation applies for  the broader $P_c(4380)$, which is experimentally found in another spin-parity channel, while  energy shifts would be larger in this case. 
 
 One of our goals is therefore to look for possible extra eigen-states in the spectrum, which could signal the presence of  pentaquarks within one-channel approximation. This underlines the importance of finding  a complete spectrum of eigenstates with all expected nearly-degenerate energy levels. Only then one could relate an extra level to a possible signature for $P_c$.   Spectra in all lattice irreducible representations are considered in our work and those include also $J^P=3/2^\pm,~5/2^\pm$ that are particularly relevant for pentaquark searches.  
 
 \vspace{0.1cm} 
 
 The paper is organized as follows. The variational approach to extracting eigen-energies is reviewed  \ref{sec:5}.   The operators for single hadrons and for nucleon-meson systems are discussed in Sections \ref{sec:2} and  \ref{sec:3}, underlying complications  arising from the fact that scattering particles $N$ and $J/\psi$ carry spin. Construction of two-hadron correlators  is described in Sections \ref{sec:4}, followed by details of the lattice simulation in Section \ref{sec:6}. Section \ref{sec:7} presents the results   and compares them to previous simulations. We end by   Conclusions and Outlook. 
 
 \section{Extracting eigen-energies }\label{sec:5}
 The aim of this work is to extract energies of the eigenstates for $NJ/\psi$ and $N\eta_c$ systems, as well as for  the relevant single hadrons $N$, $J/\psi$ and $\eta_c$. The eigen-energies of a certain system  are obtained by  computing   ${\tt N}\times {\tt N}$  correlation matrices $C$ on the lattice   
\bea
	C_{ij}(t) = \langle0| O_{i}(t)\bar{O}_{j}(0)|0 \rangle= ~~~~ ~~~~~~~~  \nonumber \\ 
	=  \sum_{n}  e^{-E_n t} \langle 0|O_i|n\rangle \langle n|\bar{O}_j |0\rangle, ~ i,j=1,...,{\tt N}~.
	\label{cij}
	\eea
where $O_{i=1,..,{\tt N}}$ denote interpolators that have quantum numbers of the desired system. 
In order to reliably extract the excited states, it is favorable to employ  a large number of relevant interpolators, like for example in \cite{hsc1}. The information on the eigenstates   is obtained by inserting complete sum of eigenstates  $|n \rangle$.      We employ the widely used variational approach 
 by solving the generalized eigenvalue problem (GEVP) \cite{Blossier2009, Luscher1990222} 
\be
C(t)u^{(n)}(t)= \lambda ^{(n)}(t)C(t_{0})u^{(n)}(t), ~ ~ n=1,...,{\tt N}
\label{eqn:GEVP}
\ee
where $t_0=2$ is employed. The eigen-energies $E_n$ are extracted from the resulting eigenvalues   
 \be
\lambda^{(n)}(t)_{\mathrm{large~ t}}=  A_n e^{-E_{n}t}\ , \quad E_{eff}^{(n)}(t)=\ln\frac{\lambda^{(n)}(t)}{\lambda^{(n)}(t+1)}
\label{eqn:lambda}
\ee
  by one-exponential fits in the plateau region.

\section{  Single-hadron operators}\label{sec:2}
\label{ch:single_hadron_interpolators}
 In order to determine the non-interacting energies of the nucleon-charmonium system, we separately compute nucleon and charmonium correlators  with momenta $p^2=0,~1,~2$ (in units of $(2\pi/{\tt L})^2)$.   Three standard  nucleon operators   	
  	\begin{align} 
		 \label{eqn:op_nucleon}
		& N(p,t)=\sum_{x}\epsilon_{abc} P^{+} \Gamma^{1} u(x,t) (u^{T}(x,t)\Gamma^{2}d(x,t)) e^{\dot{\imath} px}\nonumber \\
		& (\Gamma^{1},\Gamma^{2}):     (\mathbb{1},C\gamma_{5}) ,\ 
		(\gamma_{5},C),\ 
		(\mathbb{1},\dot{\imath} \gamma_{t}C\gamma_{5}) 
		\end{align}	
	and two standard  operators for each charmonia     
		\begin{align} \label{eqn:op_meson}
&M(p,t)=\sum_{x} c(x,t)\Gamma \bar{c}(x,t) e^{\dot{\imath} px} \\
& \Gamma(J/\psi):\   \gamma_{i} ,  \gamma_{i} \gamma_{t}, \ i=x,y,z  \nonumber\\
&\Gamma(\eta_c)\ \ :\   \gamma_{5},  \   \gamma_{t} \gamma_{5}  \nonumber
\end{align}
	are employed at each momentum $p$.

\section{ Two-hadron operators and expected degeneracies in the non-interacting case}\label{sec:3}

The operators for two-hadron system are needed to compute eigen-energies of these systems from the correlation functions.  
We consider only the system with total momentum zero since parity $P$ is a good quantum number in this case, which simplifies the study. Operators are therefore  of the form
$$ O\simeq N(p)M(-p)~,$$
where each hadron is projected to a definite momentum $p^2=0,1,2$.  
All possible quantum numbers $J^P$ are considered since $J^P$ of $P_c$ pentaquarks  are not reliably established from experiment yet. 
Therefore we consider all six irreducible representations $\Gamma^P=G_1^\pm,~G_2^\pm,~H^\pm$ of the   discrete lattice group $O_h$. 

\subsection{ Operators in partial-wave method}

The operators which have good total angular momentum $J$, its third component $m_{J}$, angular momentum $\ell$ and total spin of particles $S$ in continuum are  referred as to partial-wave (PW) operators \cite{Callat_partial_wave, operators_2017}  
\begin{eqnarray}
\label{eqn:partial_wave}
O^{|p|,J,m_J,\ell,S}=\sum_{m_\ell,m_S,m_{s1},m_{s2}} C^{Jm_J}_{\ell m_\ell,Sm_S}C^{Sm_S}_{s_1m_{s1},s_2m_{s2}}  \nonumber \\ 
\sum_{R\in O} Y^*_{\ell,m_\ell}(\widehat{Rp}) N_{m_{s1}}(Rp)M_{m_{s2}}(-Rp)~.~~~~
\end{eqnarray}
The operator has good parity $P=P_1P_2(-1)^{\ell}$. The $N\eta_c$ system carries $S=\frac{1}{2}$, while $NJ\psi
$ can have $S=\frac{1}{2}$ or $\frac{3}{2}$.  A given combination of $J^P$  can be obtained  with multiple combinations of $(S,\ell)$ and such channels can be coupled in the continuum infinite-volume (e.g. example in Eq. \ref{eqn:degeneracy_continuum}).

\begin{table}[h!]
	\begin{tabular}{c|c}
		irrep $\Gamma^P$ & $J^P$ \\
		\hline
		$G_{1}^\pm$ & $\frac{1}{2}^\pm$ ,  $\frac{7}{2}^\pm$ \\
		 $G_{2}^\pm$ & $\frac{5}{2}^\pm$ ,  $\frac{7}{2}^\pm$ \\	
		  $ H^\pm$ & $\frac{3}{2}^\pm$ , $\frac{5}{2}^\pm$ ,  $\frac{7}{2}^\pm$ \\	
	\end{tabular}
	\caption{Irreducible representations $\Gamma^P$ of the discrete lattice group $O_h$, together with a list of $J^P$ that a certain irrep contains.   }
	\label{tab:subduction_depending_on_J}
\end{table}

On the finite cubic lattice, the operators $O^{J,m_J}$ form a reducible representation with respect to the lattice group $O_h$. One has to employ operators which transform according to irreducible representations $\Gamma^P$. Those are listed in Table \ref{tab:subduction_depending_on_J}, where several $J^P$ contribute to a given irrep $\Gamma^P$. 
 Therefore partial wave operators (\ref{eqn:partial_wave}) are subduced  to chosen row $r$ of irrep $\Gamma$ 
\be
\label{eqn:subduction}
O_{|p|,\Gamma,r}^{[J,\ell,S]}=\sum_{m_J}  {\cal S}^{J,m_J}_{\Gamma,r} O^{|p|,J,m_J,\ell,S}~,
\ee 
where the  subduction coefficients $ {\cal S}^{J,m_J}_{\Gamma,r} $ are given in  the Appendix of  \cite{PolSpinEdwardsDudek2011}. 
All these operators for nucleon-vector and nucleon-pseudoscalar systems with $p^2=0,~1$ are explicitly  written   in  Appendix C of  \cite{operators_2017}. 

A simple example of the $NJ/\psi$ operator for $p=0$ that transforms according to $G_1^-$ irrep is
 \be
	O_{G_1^-,r=1}=N_{\frac{1}{2}}(0) V_z(0)+N_{-\frac{1}{2}}(0) \left(V_x(0)-\dot{\imath} V_y(0)\right)~.
	\label{eq:Op_VN_G1m}
	\ee
Let us explicitly present also a more non-trivial  example, where $N\;J/\psi$ with $p^{2}=1$ transform according to  $H^{-}$.  Here the relations (\ref{eqn:partial_wave},\ref{eqn:subduction}) lead to a number of linearly-dependent operators 
\begin{align}\label{eq:Hm_example}
&(p^2,J,\ell,S)=(1,\frac{3}{2},0,\frac{3}{2}),(1,\frac{3}{2},2,\frac{1}{2}),(1,\frac{3}{2},2,\frac{3}{2}),\nonumber\\
&\quad (1,\frac{5}{2},2,\frac{1}{2}),(1,\frac{5}{2},2,\frac{3}{2}),(1,\frac{5}{2},4,\frac{3}{2}),..
\end{align}
We find a basis of three linearly independent operators, which are chosen to be the operators in the first line. Those are listed in Table \ref{tab:qn_vm_nuk} and employed for the correlation matrices for $NJ/\psi $ system. 
The first operator, for example, has the form 
 \begin{eqnarray}
	O_{H^{-},p^{2}=1}^{(J,\ell,S)=(\frac{3}{2},0,\frac{3}{2})} =N_{\frac{1}{2}}\left(e_z\right) \left(V_x\left(-e_z\right)-\dot{\imath} V_y\left(-e_z\right)\right)  \nonumber \\
	+N_{\frac{1}{2}}\left(-e_z\right) \left(V_x\left(e_z\right)-\dot{\imath} V_y\left(e_z\right)\right)+ \nonumber \\
	N_{\frac{1}{2}}\left(e_x\right) \left(V_x\left(-e_x\right)-\dot{\imath} V_y\left(-e_x\right)\right)  \nonumber \\
	+N_{\frac{1}{2}}\left(-e_x\right) \left(V_x\left(e_x\right)-\dot{\imath} \left(e_x\right)\right)+ \nonumber \\
	N_{\frac{1}{2}}\left(e_y\right) \left(V_x\left(-e_y\right)-\dot{\imath} V_y\left(-e_y\right)\right)  \nonumber \\
	+N_{\frac{1}{2}}\left(-e_y\right) \left(V_x\left(e_y\right)-\dot{\imath} V_y\left(e_y\right)\right)
	\end{eqnarray}
Other operators are linear combinations of  three in the first line of Eq. (\ref{eq:Hm_example}), for example
\bea
O_{H^{-},p^{2}=1}^{(J,\ell,S)=(\frac{5}{2},2,\frac{1}{2})} = \frac{3}{2\sqrt{\pi}} O_{H^{-},p^{2}=1}^{(J,\ell,S)=(\frac{3}{2},2,\frac{1}{2})} \nonumber \\ 
O_{H^{-},p^{2}=1}^{(J,\ell,S)=(\frac{5}{2},2,\frac{3}{2})} =- \frac{3}{\sqrt{14 \pi}} O_{H^{-},p^{2}=1}^{(J,\ell,S)=(\frac{3}{2},2,\frac{3}{2})} \nonumber \\ 
O_{H^{-},p^{2}=1}^{(J,\ell,S)=(\frac{5}{2},4,\frac{3}{2})} =-\frac{1}{4}\sqrt{\frac{21}{\pi}} O_{H^{-},p^{2}=1}^{(J,\ell,S)=(\frac{3}{2},0,\frac{3}{2})}+ \nonumber \\ + \frac{5}{4}\sqrt{\frac{3}{7\pi}} O_{H^{-},p^{2}=1}^{(J,\ell,S)=(\frac{3}{2},2,\frac{3}{2})} ~ \nonumber 
\eea
and these are not explicitly incorporated in the correlation matrix. 

Along similar lines, we  found linearly independent  partial-wave operators for all irreducible representations and relative momenta. Our choices of  linearly-independent sets $(J,\ell,S)$ are given in  appendix   \ref{ch:appA} \footnote{Operators for other possible combinations of quantum numbers $(J,\ell,S)$ can be expressed as a linear combination of the chosen basis in Appendix  \ref{ch:appA}.}. The number of linearly independent operator-types for each $|p|$ is provided in the Table  
\ref{tab:num_of_operators}.  For each of  the operator-types listed in Appendix \ref{ch:appA} we in fact implement six operators: three  choices for the nucleon (Eq. \ref{eqn:op_nucleon}) times two choices for charmonium (Eq. \ref{eqn:op_meson}).   The final number of operators used in the computation of the correlation matrix for each irrep is given in Table \ref{tab:num_of_used_operators}.

\begin{table}[h!]
	\begin{tabular}{c||ccc|ccc}
		$\text{irrep}$ & \multicolumn{3}{c|}{$N(p)\eta_c(-p)$} & \multicolumn{3}{c}{$N(p)J/\psi(-p)$}  \\ 
		\hline 
		& $p^{2}=0$  & $p^{2}=1$  & $p^{2}=2$  & $p^{2}=0$  & $p^{2}=1$  & $p^{2}=2$  \\ 
		\hline 
		\hline 
		$G_{1}^{+}$ & $0$ & $1$ & $1$ & $0$ & $2$ & $3$ \\ 
		\hline 
		$G_{1}^{-}$& $1$ & $1$ & $1$ & $1$ & $2$ & $3$ \\ 
		\hline 
		$G_{2}^{+}$ & $0$ & $0$ & $1$ & $0$ & $1$ & $3$ \\ 
		\hline 
		$G_{2}^{-}$ & $0$ & $0$ & $1$ & $0$ & $1$ & $3$ \\ 
		\hline 
		$H^{+}$ & $0$ & $1$ & $2$ & $0$ & $3$ & $6$ \\ 
		\hline 
		$H^{-}$  & $0$ & $1$ & $2$ & $1$ & $3$ & $6$ 
	\end{tabular}
	\caption{ The number of the expected degenerate eigenstates $N(p)M(-p)$  for each row of irrep within non-interacting limit. This number is equal to the   number of the linearly-independent operator-types, which are listed in Appendix \ref{ch:appA}. }
	\label{tab:num_of_operators}
\end{table}

\begin{table}[h!]
	\begin{tabular}{c||cccccc}
$\text{irrep}$& $G_{1}^{-}$& $G_{1}^{+}$ & $G_{2}^{-}$  &$G_{2}^{+}$ & $H^{-}$& $H^{+}$ \\ 
\hline
$N\eta_{c}$
& $3 \times 6$& $2 \times 6$ & $1 \times 6$& $1 \times 6$&$3 \times 6$&$3 \times 6$\\

$NJ/\psi$ 
& $6 \times 6$&$5 \times 6$&$4 \times 6$&$4 \times 6$&$10 \times 6 $&$9 \times 6$ \\
	\end{tabular}
	\caption{Number ${\tt N}$ of interpolators used to compute the ${\tt N}\times {\tt N}$ correlation matrix $C$ (\ref{cij}) for a given row of irrep. The study of  $NJ/\psi$ scattering in $H^-$, for example, uses ${\tt N}=10 \times 6 =60$ interpolators (10 operator types with 3 nucleon and 2 vector choices for each), so   correlation matrix of size $60\times 60$   is computed. }
	\label{tab:num_of_used_operators}
\end{table}

We note that the number of linearly independent operator-types  in Table \ref{tab:num_of_operators} agrees with the number of linearly-independent operators-types obtained using the projection  method \cite{operators_2017}, helicity method \cite{operators_2017} and Clebsch-Gordan decomposition in  \cite{Moore2006,Moore2006a}.

\subsection{Expected number of degenerate states in the non-interacting limit}
\label{ch:expected_number}  

In the non-interacting limit, one expects several degenerate $N(p)M(-p)$ eigenstates  for most of $J^P$ (or irreps) and relative momenta $p\!>0\!$. Let us try to understand this first in the continuum limit of an infinite volume. 
Different combinations of $(\ell,S)$   lead to a given $J^P$ ($|\ell-S|\leq J\leq |\ell+S|$)  due to the non-zero spins of the scattering particles. An example is given in Eq. (\ref{eqn:degeneracy_continuum}), while more examples can be deduced from Tables  \ref{tab:qn_pm_nuk} and \ref{tab:qn_vm_nuk}. 
The linearly independent combinations $(\ell,S)$ represent linearly independent eigenstates, so each of them should feature as an independent eigenstate in the spectrum. 

This will remain true on the lattice on a finite volume and the corresponding discrete symmetry group.
However, in this case also  different spins $J^P$ can contribute to a given irrep $\Gamma^P$ as listed in Table \ref{tab:subduction_depending_on_J}. 
Linearly independent combinations $(J,\ell,S)$ that subduce 
 to given irrep $\Gamma^P$ now present linearly-independent eigenstates.  These are listed in appendix  \ref{ch:appA} and their number is summarized in Table \ref{tab:num_of_operators} for each $\Gamma^P$ and relative momentum $p^2$. In the non-interacting limit, each of those linearly independent eigenstates should appear in the spectrum. Therefore  the expected number of degenerate eigenstates for each row  is given in Table \ref{tab:num_of_operators}.  
These degeneracies can be lifted by the presence of the  interactions.

\section{Construction of two-hadron correlators}  \label{sec:4}
\label{ch:combining}

\begin{figure}[h!]
	\centering
	\includegraphics[width=0.18\textwidth]{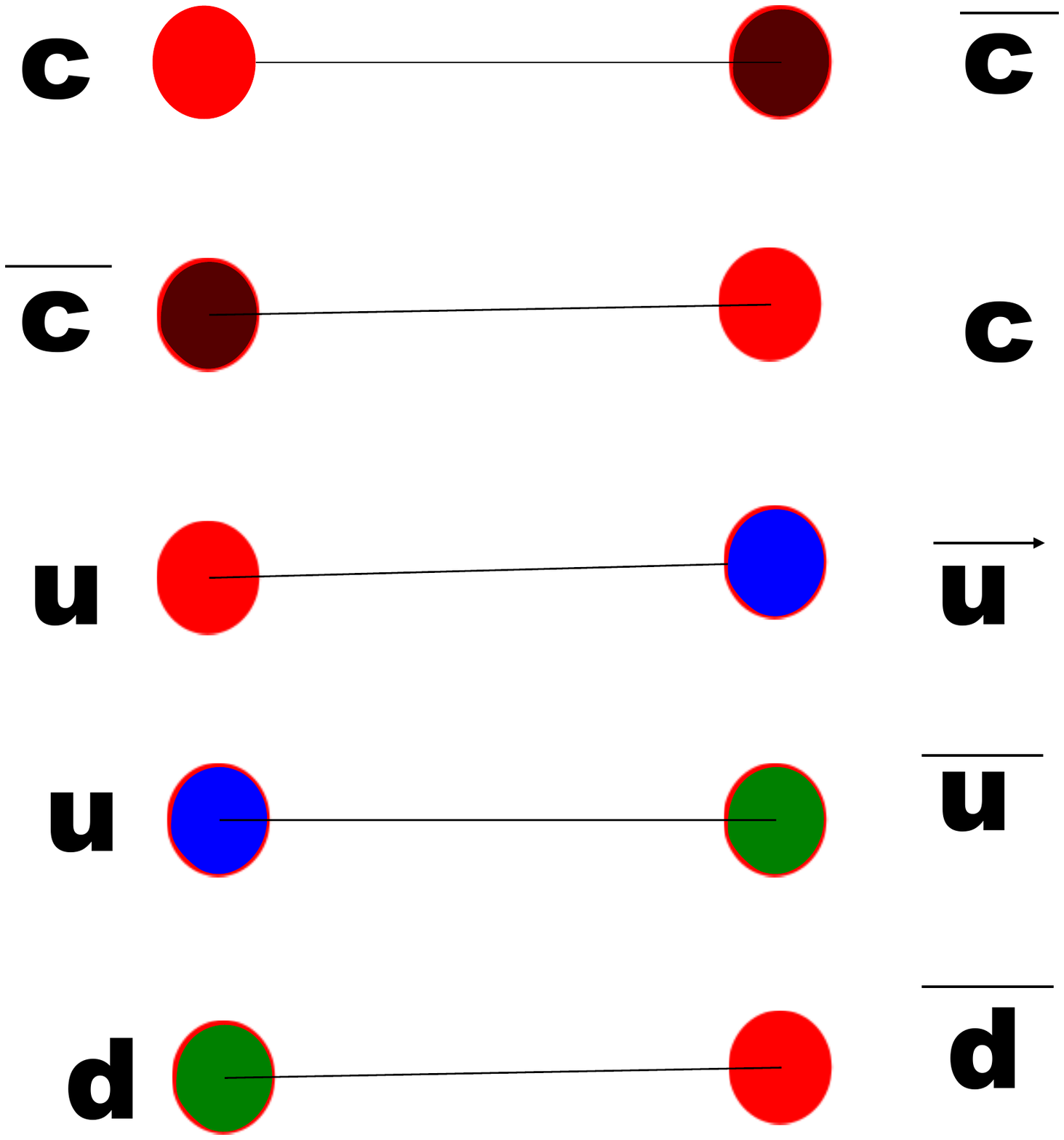}
	\includegraphics[width=0.18\textwidth]{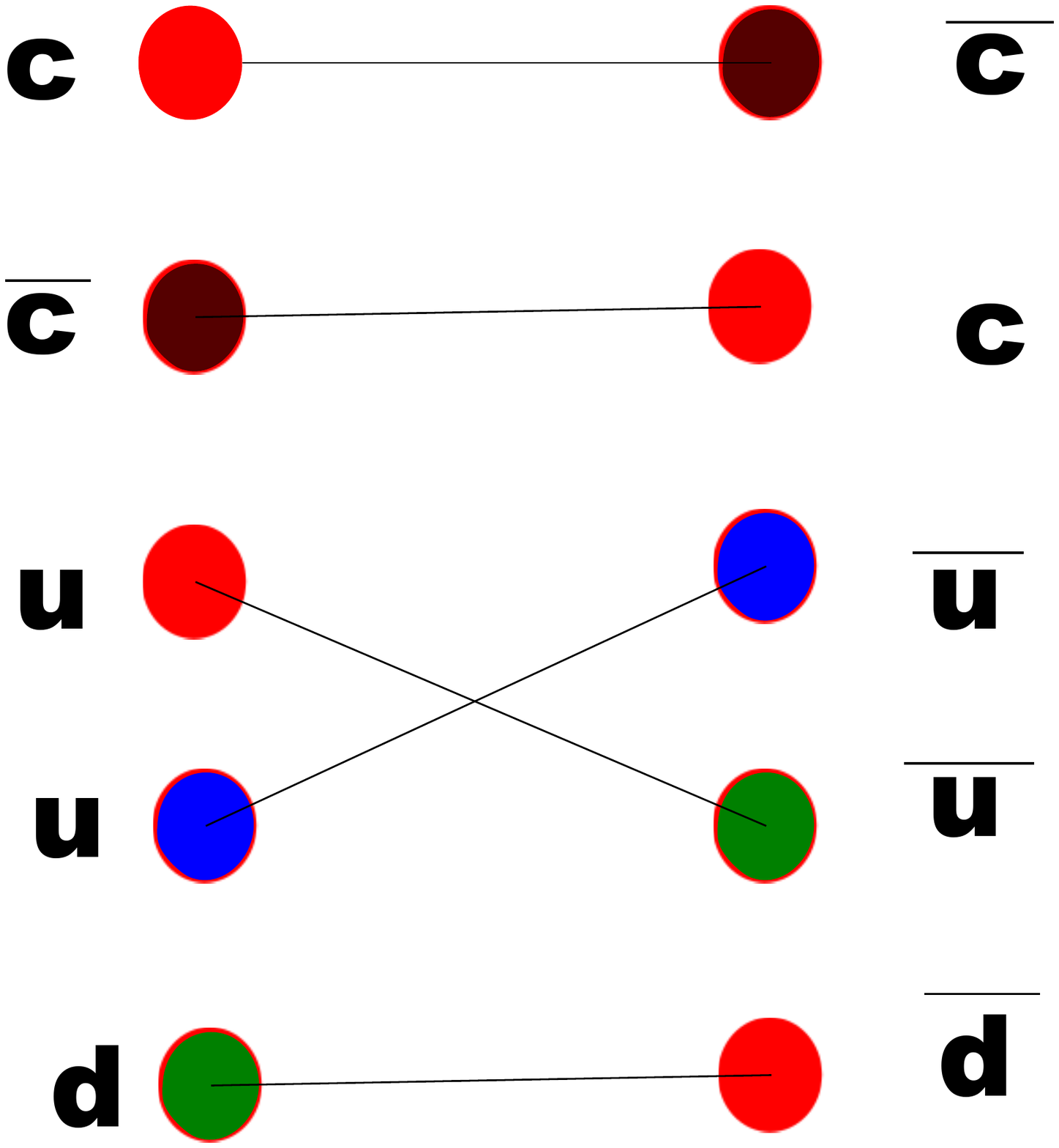}
	\caption{Wick contractions considered in our simulation for one-channel  approximation.}
	\label{fig:pccontr}
\end{figure}

In the one-channel approximation there is no contraction connecting charmonium meson and nucleon interpolator, as shown in Figure \ref{fig:pccontr}.   
Therefore single hadron correlation functions  can be simulated separately and then later combined to the two-hadron correlation function.   All   two-hadron correlators in our study can be expressed in terms of 
\begin{align}\label{two-point-cors}
 &\langle0|N_{m_s'}(p')\bar{N}_{m_s}(p)|0\rangle_c,~~\\
  &\langle0|V_{i'}(p')V_{i}^\dagger (p)|0\rangle_c,~~ \langle0|P(p') P^\dagger(p)|0\rangle_c, ~~ V=J/\psi, ~ P=\eta_c\nonumber
\end{align}
 which are pre-computed for all combinations of $p',p=0,1,2$, $i,i'=x,y,z$ and $m_s,m_s^\prime=1/2,-1/2$, as well as for all configurations $c$.  We omit the Wick contractions with the disconnected charm contributions in the study of pentaquark as well as charmonium systems. These contractions induce charmonium decays to the light hadrons and have been omitted in most of the previous lattice studies related to charmonium-like systems. 

Let us give an example of two-hadron correlator corresponding to operator $O_{G_1^-,r=1}$   (Eq. \ref{eq:Op_VN_G1m}) at the sink and  conjugate operator  
\be
	\bar{O}_{G_1^-,r=1}=\bar{N}_{\frac{1}{2}}(0) \bar{V}_z(0)+\bar{N}_{-\frac{1}{2}}(0) \left(\bar{V}_x(0)+\dot{\imath} \bar{V}_y(0)\right)
	\ee
at the source.    Given that  there is no contractions connecting $J/\psi $ and $N$ interpolators (Fig. \ref{fig:pccontr}),   the two-hadron correlator can be expressed as 
 \begin{align}
	&\langle0|O_{G_1^-,r=1} \bar{O}_{G_1^-,r=1}|0\rangle= ~~~~~~~~~ \nonumber \\
	~~~ \Big\langle ~&\langle0|N_{\frac{1}{2}}(0)\bar{N}_{\frac{1}{2}}(0)|0\rangle_c\langle0|V_z(0) \bar{V}_z(0)|0\rangle_c+  ~~~~ \nonumber \\
	+&\langle0|N_{-\frac{1}{2}}(0)\bar{N}_{\frac{1}{2}}(0)|0\rangle_c\langle0|V_x(0) \bar{V}_z(0)|0\rangle_c-  ~~~~\nonumber \\
	- \dot{\imath}  &\langle0|N_{-\frac{1}{2}}(0)\bar{N}_{\frac{1}{2}}(0)|0\rangle_c\langle0|V_y(0) \bar{V}_z(0)|0\rangle_c+  ~~~~ \nonumber \\
	+ &\langle0|N_{\frac{1}{2}}(0)\bar{N}_{-\frac{1}{2}}(0)|0\rangle_c\langle0|V_z(0) \bar{V}_x(0)|0\rangle_c+  ~~~~ \nonumber \\
	+&\langle0|N_{-\frac{1}{2}}(0)\bar{N}_{-\frac{1}{2}}(0)|0\rangle_c\langle0|V_x(0) \bar{V}_x(0)|0\rangle_c-  ~~~~ \nonumber \\
	-\dot{\imath} &\langle0|N_{-\frac{1}{2}}(0)\bar{N}_{-\frac{1}{2}}(0)|0\rangle_c\langle0|V_y(0) \bar{V}_x(0)|0\rangle_c+  ~~~~ \nonumber \\
	+\dot{\imath}&\langle0| N_{\frac{1}{2}}(0)\bar{N}_{-\frac{1}{2}}(0)|0\rangle_c\langle0|V_z(0) \bar{V}_y(0)|0\rangle_c +  ~~~~ \nonumber \\
	+\dot{\imath} & \langle0|N_{-\frac{1}{2}}(0)\bar{N}_{-\frac{1}{2}}(0)|0\rangle_c\langle0|V_x(0) \bar{V}_y(0)|0\rangle_c+  ~~~~ \nonumber \\
	+ &\langle0|N_{-\frac{1}{2}}(0)\bar{N}_{-\frac{1}{2}}(0)|0\rangle_c\langle0|V_y(0) \bar{V}_y(0)|0\rangle_c~\Big\rangle~. \label{combination}
	\end{align}
The two-hadron correlation function is a  sum of products of pre-calculated single-hadron correlators. Products of nucleon and vector correlators, that are calculated on individual configurations $c$, are averaged over configurations, as indicated by the large parentheses $\langle ... \rangle$.

All ${\tt N}\times {\tt N}$ correlation functions  were constructed in a similar manner, where the number of interpolators $n$ is given in the Table \ref{tab:num_of_used_operators}.  We calculated the correlation functions for all rows of irreps
 and then averaged them over the rows to gain better statistical accuracy.

\section{Lattice setup}\label{sec:6}

\begin{table}[h!]	\begin{tabular}{c c c c c c }
		$N_{L}^{3} \times N_{T}$ & $\beta$ & $a\text{[fm]}$ & ${\tt L}\text{[fm]}$  & $\# \text{config}$ & $m_{\pi}\text{[MeV]}$ \\
		\hline
		$16^{3} \times 32$ & $7.1$ & $0.1239(13)$ & $1.98(2)$  & $281$ & $266(3)$ 
	\end{tabular}
	\caption{ Parameters of the lattice ensemble. }
	\label{tab:latt}
\end{table}

The simulation is performed on the $N_f=2$ ensemble with parameters listed in table \ref{tab:latt},  that was generated in context of the work   \cite{Latt_data_Hasenfratz_2008,Latt_data_Hasenfratz_2_2008}.  Wilson Clover action was used for light $u/d$ quarks, while  Fermilab approach \cite{fermilab} was employe for the charmed quarks.  Strange quark is not dynamical in this simulation;  we expect that dynamical strange quark is not crucial for this channel which does not involve strange valence quarks.  Decay channels to hadrons with strange quarks also do not feature 
as important channels in the phenomenological studies of $P_c$. 
Further details about the ensembles and treatment of the charm quark may be found in \cite{Latt_data_Hasenfratz_2008,Latt_data_Hasenfratz_2_2008,LangVenduciNpi,charmonium2013}.  Periodic boundary conditions in space and anti-periodic boundary conditions in time are employed for fermions. The small spatial size (in comparison to other currently used lattices) $L \approx 2 ~\text{fm}$ of our lattice has a practical advantage since only levels up to relative momentum  $p^2\leq 2 ~(2\pi/{\tt L})^2$ have to be extracted to cover the $P_c$ resonance region. A larger spatial size would imply denser energies (\ref{eqn:non_intracting_energy}) and require extracting a larger number of energy states, which would be more challenging in view of  degeneracies appearing in the spectra.

The quark fields are smeared according to the full distillation method \cite{distillation_Peardon2009}, which  allows the calculation of all the necessary correlation matrices.  The smearing of the light quarks in the nucleon is based on $N_v=48$ lowest Laplace eigenvectors $v^k$ where $k=1,..,N_v$. The smearing of the charmed quarks in the charmonium is based on $96$ eigenvectors. 
Perambulator $\tau^{k^\prime k}(t^\prime,t)$  represents  the quark propagator from the Laplace eigenvector $k$ at time $t$ to the Laplace eigenvector $k^\prime$ at time $t^\prime$. All light-quark and charmed-quark perambulators are precomputed and saved for all distillation indices $k,k^\prime=1,..,N_v$ and $t,t^\prime=1,..,N_T$. From those we compute charmonium two-point functions and nucleon two-point functions (\ref{two-point-cors}) needed for this study.  Charmonium two-point function is given as a product to two perambulators in Eq. (11) of \cite{distillation_Peardon2009}, while nucleon two-point function is a product of three perambulators in Eq. (19) of \cite{distillation_Peardon2009}. The numerical cost of computing these correlators from given perambulators comes from summing over the distillation indices ($k,k^\prime,..$). The cost is  considerable for the nucleon two-point functions, which have to be evaluated for all combinations of momenta and polarizations at source and sink  (\ref{two-point-cors}).

In the Fermilab approach to the discretization of charm quarks \cite{fermilab}, one can attribute physical significance only to the quantities where the rest energy of the charm quarks cancels.  This cancels, for example,  in the difference between the energy of the $\bar ccuud$ system and the charmonium $\bar cc$. We choose the mass of the spin-averaged $1S$ charmonium
\begin{equation}\label{eref}
m_{\bar cc,sa}\equiv \tfrac{1}{4}(m_{\eta_c}+3~m_{J/\psi})
\end{equation}
as the reference energy since this mass  is least prone to the discretization errors.  We will  compare the lattice energies   $E^{lat}-m_{\bar cc,sa}^{lat}$ to the experimental masses     $m_{P_c}^{exp}-m_{\bar cc,sa}^{exp}$.  

\section{Results}\label{sec:7}

 The resulting eigen-energies of the single hadrons    ($N$, $J/\psi$, $\eta_c$) and  two hadron systems ($NJ/\psi$, $N\eta_c$)  are presented in this  Section. They are obtained from the correlated one-exponential fits (\ref{eqn:lambda}) and the errors are  determined using jack-knife method.  
 
\begin{table}[h!]
	\begin{tabular}{c c | c c   c}
		\text{particle} &$p^{2}$ & $E_H(p)a$   &  $\sigma_{E}a$ & \text{fit range} \\
		\hline
		\hline
		$N$ &  $0$& $0.701$ & $0.019$ &  $[6,9]$ \\ 
	&  $1$& $0.769$ & $0.028$& $[7,10]$ \\ 
& $2$& $0.849$ & $0.054$ & $[7,9]$ \\ 
		\hline
		$J/\psi$& 	$0$& $1.539$ & $0.001$ & $[10,14]$ \\ 
	& 	$1$& $1.576$ & $0.001$ &$[10,14]$ \\ 
& 	 	$2$& $1.613$ & $0.001$ & $[9,12]$ \\ 
		\hline
		$\eta_{c}$ & $0$& $1.472$ & $0.001$ & $[9,12]$ \\ 
	&$1$& $1.515$ & $0.001$ & $[6,12]$ \\ 
	 & $2$& $1.551$ & $0.001$ & $[9,11]$ \\ 
	\end{tabular}
	\caption{  Energies $E_H(p)$ and their errors $\sigma_E$ for single hadrons with $p^2=0,1,2$ in units of $(2\pi/{\tt L})^2$.  }
	\label{tab:single_hadron_fit}
\end{table}

\subsection{Single hadron energies}
\label{ch:single_hadron_results}
The single-hadron energies $E_H(p)$ on the employed lattice are needed  to determine the non-interacting  energies of two hadrons (\ref{eqn:non_intracting_energy}). The fitted energies for various momenta are collected in Table \ref{tab:single_hadron_fit}.  They arise from $2\times 2$ correlation matrices where both meson operators (Eq. \ref{eqn:op_meson}) are used, while the first and the third operators (Eq. \ref{eqn:op_nucleon})  are used for the nucleon. Results for $E_H$ on each re-sampled jack-knife will be used to determine $E^{n.i}$ and energy-shifts $\Delta E$ (\ref{eqn:non_intracting_energy}) in the next subsection.

\begin{table}[h!]
	\begin{tabular}{c||cccccc}
$\text{irrep}$& $G_{1}^{-}$& $G_{1}^{+}$ & $G_{2}^{-}$  &$G_{2}^{+}$ & $H^{-}$& $H^{+}$ \\ 
\hline
$N\eta_{c}$
& $3 \times 2$& $2 \times 2$ & $1\times 2$& $1 \times 2$&$3 \times 2$&$3 \times 2$\\

$NJ/\psi$ 
& $6 \times 2$&$5 \times 2$&$4 \times 2$&$4 \times 2$&$10 \times 2 $&$9\times 2$ \\
	\end{tabular}
	\caption{Number  of interpolators ${\tt N}$ employed in the final analysis of the correlation matrix (\ref{cij}) for each irrep. The final analysis of $NJ/\psi$ scattering in  $H^-$, for example, employs ${\tt N}=10 \times 2 =20$ interpolators (10 operator types with  2 vector choices for each), so  the correlation matrix of size $20\times 20$    is analyzed. }
	\label{tab:num_of_analysed_operators}
\end{table}

\subsection{Two hadron energies}
\label{ch:scattering_results}
In the non-interacting limit, the two-hadron energies are equal to the sum of energies of the individual hadrons (Eq. \ref{eqn:non_intracting_energy}) given in Table \ref{tab:single_hadron_fit}.  
Resonances in LQCD  manifest themselves by the non-zero energy shifts  with respect to the non-interacting ones \cite{Luscher1986,Luscher1990222,Luscher1991531,Luscher:1991cf,Briceno_2014}. Therefore we calculate the energy  spectrum for all irreducible representations $\Gamma^P$ of the lattice group $O_{h}$, which contain contributions from various $J^P$ as indicated in Table \ref{tab:subduction_depending_on_J}.    Channels with possible $P_c$ candidates  are contained in  the irreps $G_{2}^{\pm}$ for $J^{P}=\frac{5}{2}^{\pm}$ and in irreps $H^{\pm}$ for $J^{P}=\frac{5}{2}^{\pm} \text{ or } \frac{3}{2}^{\pm}$.  

Eigen energies are calculated from the correlation matrices as described in Section \ref{ch:combining}.  The size of the calculated correlation matrices for all irreps is displayed in Table  \ref{tab:num_of_used_operators}. These large correlation matrices render  rather noisy eigenvalues, therefore we restricted our analysis to a somewhat smaller subset in Table \ref{tab:num_of_analysed_operators}, where each operator type (listed in Appendix \ref{ch:appA}) is represented by two meson operators (\ref{eqn:op_meson}) and  the first nucleon operator from (\ref{eqn:op_nucleon}). 
All resulting eigen-energies are obtained from the correlated one-exponential fits of the eigenvalues in the time range $t=[7,10]$ \footnote{Two-exponential fits starting from earlier time-slices lead to compatible results. In few cases these are less stable and have larger errors.}. The energies of the $\bar ccuud$ system will be provided with respect to the spin-averaged charmonium mass $(m_{\eta_c}+3m_{J/\psi})/4$, where the rest energies of the valence charm quark-pair cancels.

  \begin{figure}
	\centering
	\includegraphics[width=0.9\linewidth]{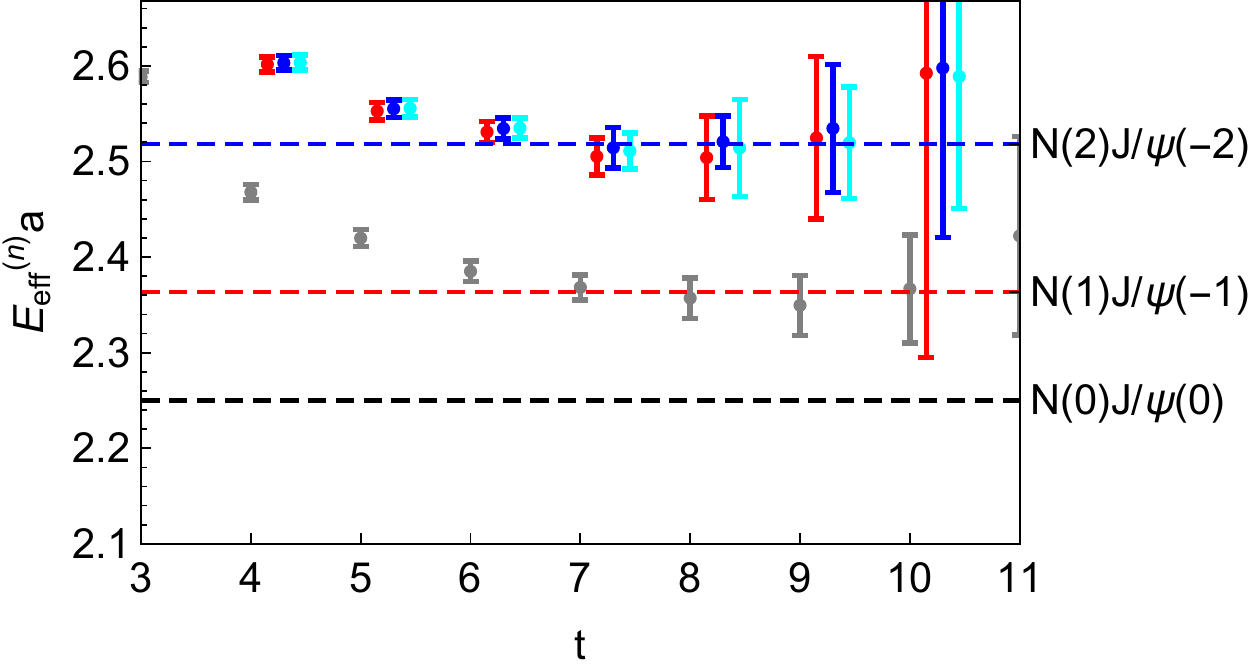}
	\caption{ Effective energies (\ref{eqn:lambda}) for the lowest four eigenstates of the   $NJ/\psi$ system in $G_2^{+}$ irrep. This gives the eigen-energy $E_n$  in the plateau region. 
	 We observe all $N(p)J/\psi(-p)$ eigenstates, expected in the non-interacting limit: this number is     $0,~1$ and $3$ states for $p^2=0,~1$ and $2$, respectively (Table \ref{tab:num_of_operators}). No additional eigenstate is found.  The non-interacting energies  (\ref{eqn:non_intracting_energy}) are indicated  by the dashed lines. Color-coding of different eigenstates is arbitrary.}
	\label{fig:effmirrepg2pbasis110000110000110000110000}
\end{figure}

\subsubsection{$NJ/\psi$ channel}
The charmed $P_{c}$ resonances were observed as two peaks in the spectrum of proton$-J/\psi$ invariant masses. Therefore, this channel  could be the promising one in which to look for the charmed pentaquarks. 

\begin{figure*}[p]
	\centering
	\includegraphics[width=0.9\linewidth]{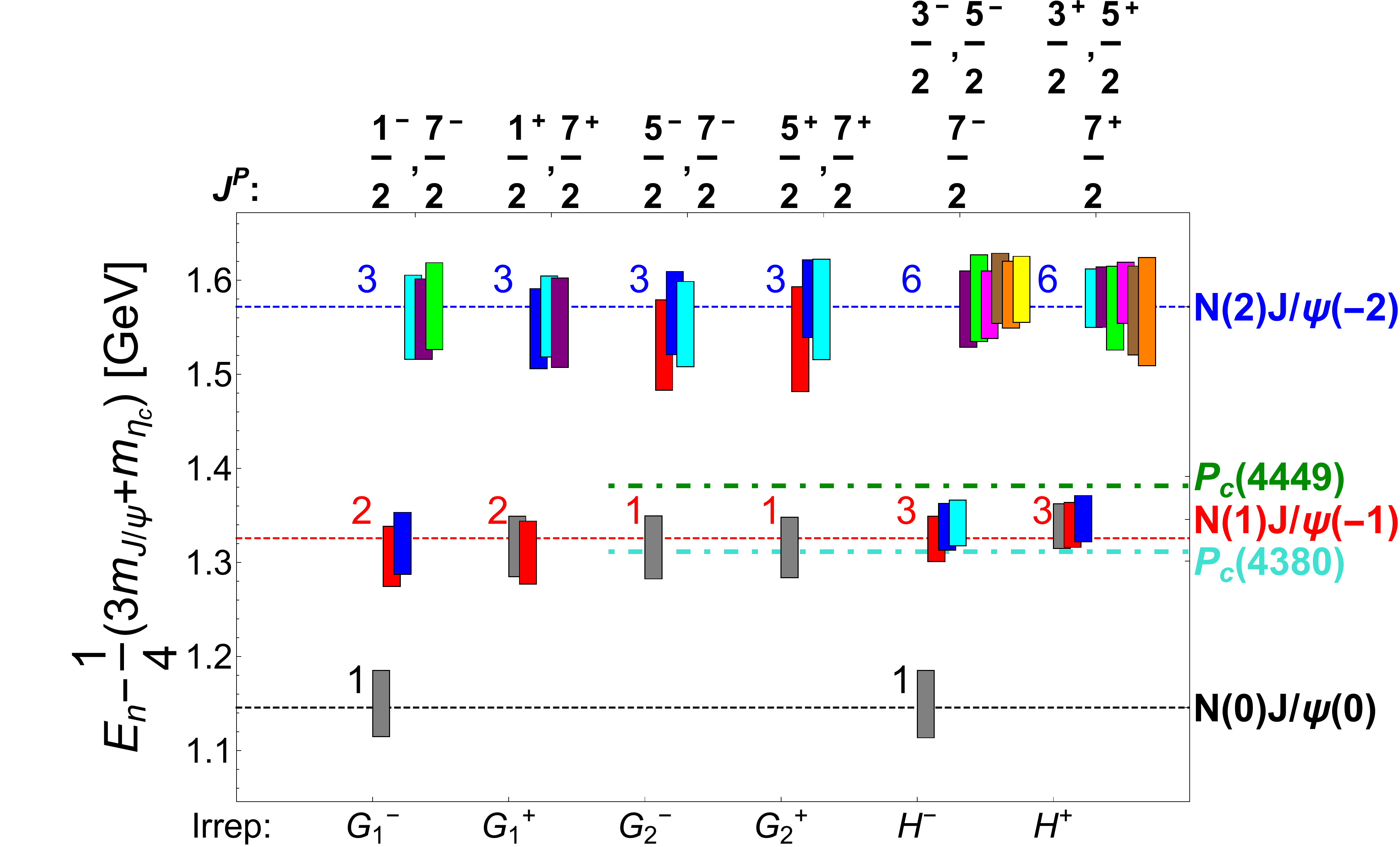}
	\caption{Energies of $NJ/\psi$ eigenstates in the one-channel approximation for all lattice irreducible representations. The quantum numbers $J^P$ that contribute to each irrep are listed on the top. Each box represents one eigenstate. The centre of the box represents it energy $E_n$, while height represents to $2\sigma_{E_{n}}$.   Ten lowest eigen-energies are, for example, shown in the irreducible representation $H^-$. The number of the observed near-degenerate states agrees with the  expected number of states in the non interacting  limit.  This number is  given in  Table \ref{tab:num_of_operators} and indicated in the plot; it arises due to the  non-zero spins of the nucleon and $J/\psi$.    Dashed lines represent  non-interacting energies $E_{N}(p) + E_{J/\psi}(-p) $   for different value of relative momentum $p$: black for $p^2=0$, red for $p^2=1$ and blue for $p^2=2$ where $p^2$ is given in units of $(2\pi/{\tt L})^2$. The dash-dotted (green and turquoise) lines correspond to experimental masses of $P_{c}$ states. The observed spectrum shows no significant energy shifts or additional eigenstates. Energies are presented with a respect to spin-averaged charmonium mass $(m_{\eta_c}+3 m_{J/\psi})/4$ (\ref{eref}). Color-coding of the eigen-energies is arbitrary.} 
	\label{fig:vm_nuk_energy_spectrum}
	\centering
	\includegraphics[width=0.6\linewidth]{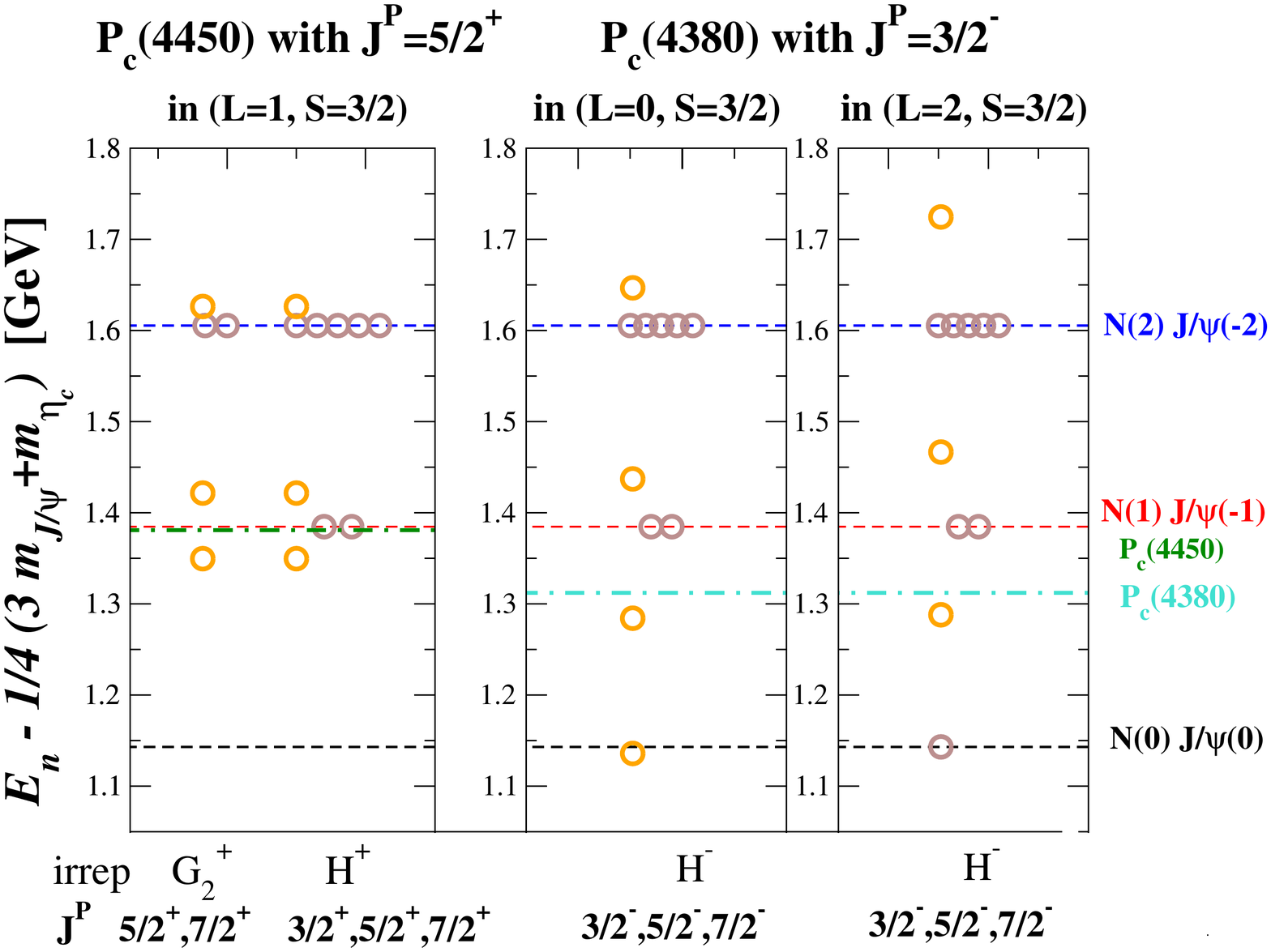}
	\caption{The energies of eigenstates in a scenario with a Breit-Wigner-type $P_c(4450)$ or $P_c(4380)$ resonances, assuming that they are coupled only to $NJ/\psi$ channel and decoupled from other two-hadron channels. This scenario renders an additional eigenstates near  $M_{P_c}\pm \Gamma_{P_c}$ with respect to the  non-interacting case.   The experimental masses $M_{P_c}$ are indicated by dashed-dotted lines. $P_c$ with the favoured $J^P$   \cite{LHCb_Pc_2015} are considered, where $\frac{5}{2}^+$ contributes to irreps $G_2^+$ and $H^+$, while $\frac{3}{2}^-$ contributes to $H^-$. The $P_c$ is assumed to reside in a single partial wave $(\ell,S)$, indicated above the plot.Dashed lines represent  non-interacting energies $E_{N}(p) + E_{J/\psi}(-p) $   for different values of relative momenta $p$. Brown circles denote levels that are not shifted with respect to non-interacting energies, while orange circles denote  levels that are shifted. }
	\label{fig:Pc_analytic}
\end{figure*}

The final eigen-energies of $NJ/\psi$ system   are presented in Figure \ref{fig:vm_nuk_energy_spectrum} and in Table \ref{tab:vm_nuk_fit} for all six irreducible representations.  An example of the effective energies   in  irrep $G_2^+$ is given in Figure \ref{fig:effmirrepg2pbasis110000110000110000110000}. The  energies of the states that are not plotted are higher and have typically (much) larger error-bars.
Let us compare this  spectrum separately with the   non-interacting limit and  with a scenario featuring $P_c$. 

\begin{itemize}
\item
 First we compare the resulting energies in Fig. \ref{fig:vm_nuk_energy_spectrum} with the {\it expectation from the non-interacting limit}. The non-interacting energies of $N(p)J/\psi(-p)$ (Eq. \ref{eqn:non_intracting_energy}) are given by the horizontal dashed lines: these energies are obtained as a sum separate lattice energies of $N$ and $J/\psi$ given in Table  \ref{tab:single_hadron_fit}. The observed energies  in Fig. \ref{fig:vm_nuk_energy_spectrum} agree with the non-interacting energies within errors.   The number of the expected degenerate  eigenstates  in the non-interacting limit is listed in  Table \ref{tab:num_of_operators}  and in Fig.  \ref{fig:vm_nuk_energy_spectrum}. Such multiplicities of levels arise due to the non-zero spin of the scattering particles $N$ and $J/\psi$.     The energies in Fig.   \ref{fig:vm_nuk_energy_spectrum} indicate that we observe exactly the same pattern of degeneracy as in the non-interacting limit. The use of carefully-constructed interpolators  \cite{operators_2017,Callat_partial_wave} were crucial for this.   We find no extra eigenstates in addition to those expected in the non-interacting limit. So, the observed  lattice spectrum is rather close to the non-interacting case. 
\item
Next we compare  the  energies   with  the {\it analytic prediction based on the existence of 
$P_c(4450)$ or $P_c(4380)$} in Fig. \ref{fig:Pc_analytic}. The aim is to explore the one-channel scenario where $P_c$ is coupled to $NJ/\psi$ and decoupled from other two-hadron decay channels.  
Such scenario renders an additional eigenstates near  $M_{P_c}\pm \Gamma_{P_c}$ with respect to the  non-interacting case (Fig. \ref{fig:Pc_analytic}).  
The favored channels $J^P=5/2^+$ for $P_c(4450)$ and $3/2^-$ for $P_c(4380)$ \cite{LHCb_Pc_2015} are considered\footnote{The same conclusions apply for other possible $J^P$ listed in the introduction.}. We assume that $P_c$  resides only in a single partial wave $(\ell,S)$ and  that there is no interaction in the other channels.  The  Breit-Wigner-type  dependence of the phase shift is employed, with the resonance parameters $M_{P_c}$ and $\Gamma_{P_c}$ taken from experiment \cite{LHCb_Pc_2015}\footnote{The resonance mass in (\ref{eq:Pc_analytic}) is taken to be $M_{P_c}^{exp}-m_{s.a.}^{exp}+m_{s.a.}^{lat}$ with $m_{s.a.}=(3m_{J/\psi}+m_{\eta_c})/4$, in accordance with Fig.   \ref{fig:vm_nuk_energy_spectrum}. }   
\begin{align}\label{eq:Pc_analytic}
&\cot \delta_{(\ell,S)}=\frac{M_{P_c}-E^2}{E~\Gamma(E)}= \frac{2Z_{00}(1;~p^2(\tfrac{2\pi}{\tt L})^2)}{\sqrt{\pi}~{\tt L}~p}\\
&\Gamma(E)=\Gamma_{P_c} \biggl(\frac{p(E)}{p(M_{P_c})}\biggr)^{2\ell+1}\frac{M_{P_c}^2}{E^2}~.\nonumber
\end{align}
The relation between the eigen-energies $E$ and the phase shift $\delta$  for the scattering of particles with arbitrary spin was derived in  \cite{Briceno_2014}. This reduces to the well-known L\"uschers relation  \cite{Luscher1986,Luscher1990222,Luscher:1991cf,Luscher1991531}    
for a resonance that appears only in the channel $(J^P,\ell,S)$ (\ref{eq:Pc_analytic}). The analytic predictions for the energies (orange circles) are obtained by solving Eq. (\ref{eq:Pc_analytic})  for $E$, assuming the continuum dispersion relation $E(p)=(p^2+m_N^2)^{1/2}+(p^2+m_{J/\psi}^2)^{1/2}$ and employing  $m_{N,J/\psi}$ determined from the lattice.  Other levels in a given irrep  remain intact and have non-interacting energies (brown circles).  The predictions  in  Fig. \ref{fig:Pc_analytic}  show all possible irreps and choices of partial waves $\ell \leq 2$ where the $P_c$ resonance with given $J^P$ could reside. 
 
    The scenario with a $P_c$ resonance in Fig. \ref{fig:Pc_analytic}  features an additional eigenstate (with respect to the non-interacting case) in the energy range roughly $M_{P_c}\pm \Gamma_{P_c}$, while some of the other levels near the resonance region get shifted. 
    So, this scenario predicts one  eigen-state more (with respect to Table \ref{tab:num_of_operators})   in the explored energy region within  the corresponding irreducible representation. We do not observe such an additional eigenstate, so such scenario featuring   $P_c$ is not supported by our lattice data. 
         \end{itemize}        
         In summary, our lattice spectra  in Fig. \ref{fig:vm_nuk_energy_spectrum} are roughly in agreement with the predictions for almost non-interacting $N$ and $J/\psi$. These spectra do not support the  scenario where a $P_c$ resonance couples only to $N/\psi$ decay channel and is decoupled from other channels. These results indicate that the  existence of $P_c$ resonance  within a one-channel   
         $NJ/\psi$ scattering  is not favored in QCD. This might suggest that the strong coupling between the $NJ/\psi$ with other channels might be responsible for the existence of the $P_c$ resonances in experiment.  Future lattice simulations  of the coupled-channel scattering will be needed to confirm or refute this hypothesis.

\subsubsection{$N\eta_{c}$  channel}

\begin{figure}[h!]
	\centering
	\includegraphics[width=0.8\linewidth]{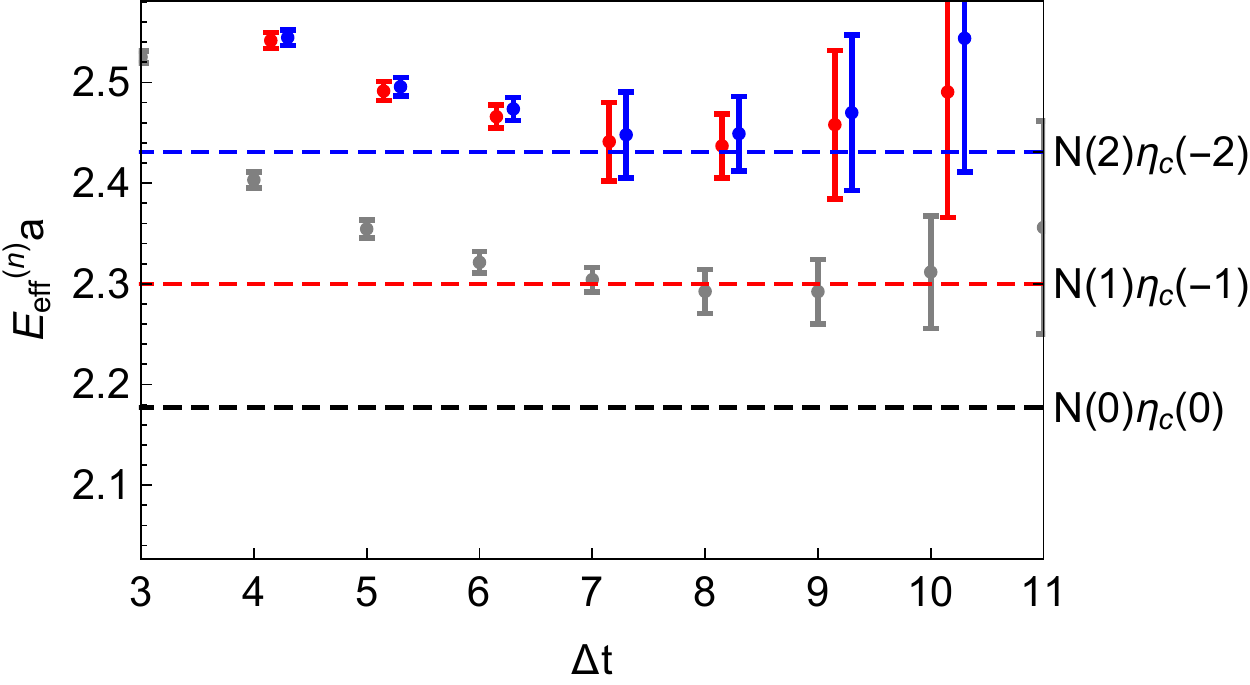}
	\caption{Effective energies for  $N\eta_{c}$ system in $H^{-}$ irrep.
	 We observe all $N(p)\eta_c(-p)$ eigenstates expected in the non-interacting limit: this number is     $0,~1$ and $2$ states for $p^2=0,~1$ and $2$, respectively (Table \ref{tab:num_of_operators}). No additional eigenstate is found.  The non-interacting energies (\ref{eqn:non_intracting_energy}) are indicated  by the dashed lines. Color-coding of eigen-energies is arbitrary. }
	\label{fig:effmirrephmbasis110000110000110000}
\end{figure}

\begin{figure*}
	\centering
	\includegraphics[width=0.85\linewidth]{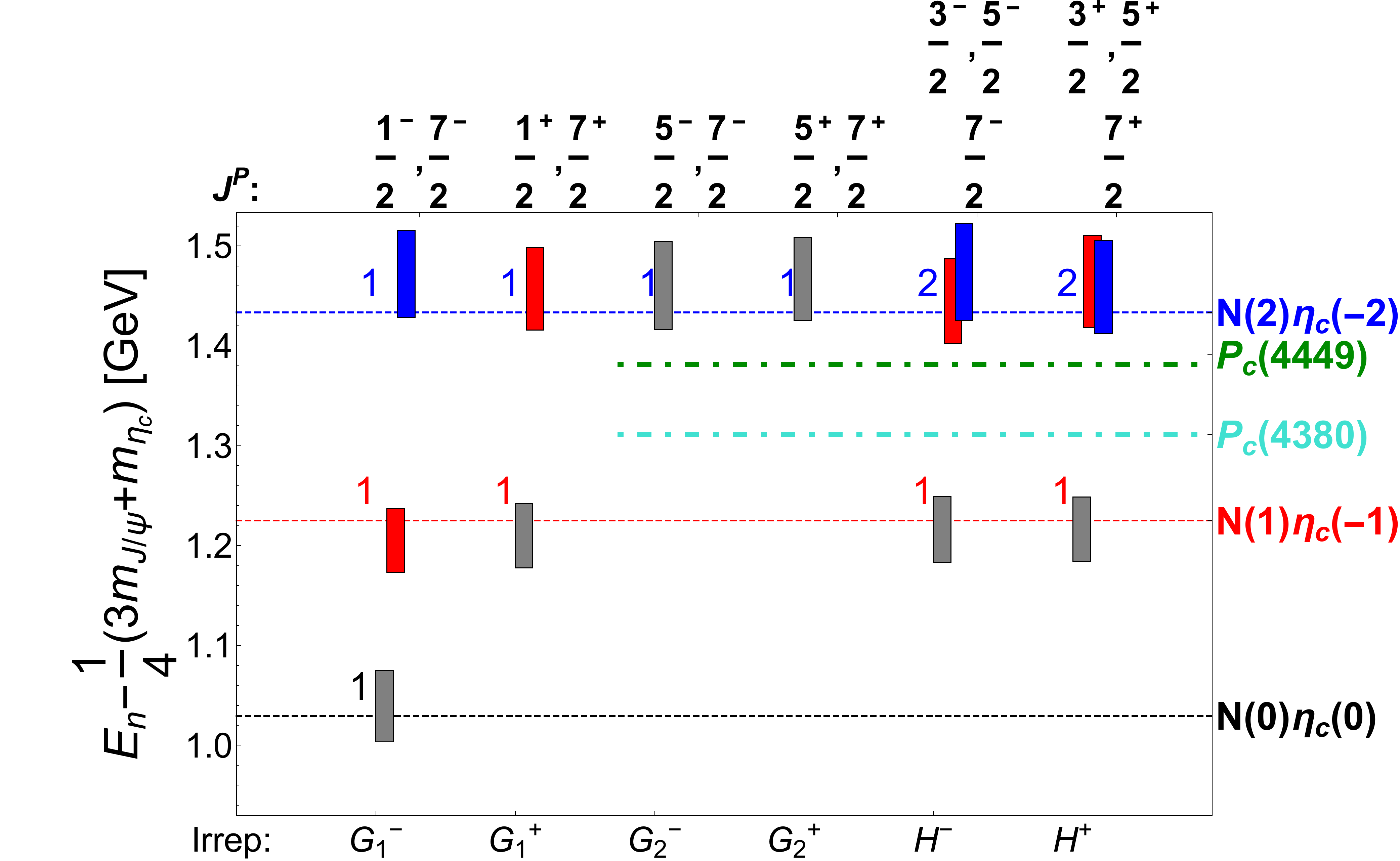}
	\caption{Energies of $N\eta_{c}$ eigenstates  in the one-channel approximation for all lattice irreducible representations. 
	The quantum numbers $J^P$ that contribute to each irrep are listed on the top. Each box represents one eigenstate. The centre of the box represents it energy $E_n$, while height represents to $2\sigma_{E_{n}}$.   The number of the observed near-degenerate states agrees with the  expected number of states in the non interacting  limit.  This number is  given in  Table \ref{tab:num_of_operators} and indicated in the plot; it 
	arises due to the  non-zero spin   of the nucleon.  This number is smaller for $N\eta_c$ than for $NJ/\psi$ since $J/\psi$ carries spin one, while $\eta_c$ is spinless.   Dashed lines represent  non-interacting energies $E_{N}(p) + E_{\eta_c}(-p) $ (Eq. \ref{eqn:non_intracting_energy})  for different values of the relative momenta $p$: black for $p^2=0$, red for $p^2=1$ and blue for $p^2=2$, where $p^2$ is given in units of $(2\pi/{\tt L})^2$. The dash-dotted (green and turquoise)   lines correspond to experimental masses of $P_{c}$ states. The observed spectrum shows no significant energy shifts or additional eigenstates.   Energies are presented with respect to spin-averaged charmonium mass $(m_{\eta_c}+3 m_{J/\psi})/4$ (\ref{eref}).}
	\label{fig:pm_nuk_energy_spectrum}
\end{figure*}

The final eigenenergies of $N\eta_c$ system   are presented in Figure \ref{fig:pm_nuk_energy_spectrum} and Table \ref{tab:pm_nuk_fit} for all six irreducible representations.  An example of effective energies in  irrep $H^-$ is given in Fig. \ref{fig:effmirrephmbasis110000110000110000}. 

\begin{itemize}
\item
First we compare the resulting energies in Fig. \ref{fig:pm_nuk_energy_spectrum} with the {\it expectation from the non-interacting limit}. The non-interacting energies of $N(p)\eta_c(-p)$ (Eq. \ref{eqn:non_intracting_energy}) are indicated by the horizontal dashed lines. We find that the observed energies  agree with the non-interacting $N\eta_c$ within sizable errors of our calculation.   The number of expected degenerate  eigenstates  in the non-interacting limit is listed in  Table \ref{tab:num_of_operators}  and in Fig. \ref{fig:pm_nuk_energy_spectrum}. The degree of degeneracy is smaller than in $NJ/\psi$ case since $\eta_c$ does not carry spin. We observe exactly the same pattern of degeneracy, while no additional eigenstates are found.
   \item
 The {\it scenario featuring $P_c$}, coupled dominantly to $N\eta_c$ and largely decoupled from other channels\footnote{In reality, $P_c$ can  not be   coupled only to $N\eta_c$, as it was experimentally observed in $NJ/\psi$ decay.}, would predict an additional eigenstate in the energy range roughly $M_{P_c}\pm \Gamma_{P_c}$.  This is based  on an analogous argument, presented in more detail  for $NJ/\psi$ system.  Such an additional eigenstate is not observed in Fig.  \ref{fig:pm_nuk_energy_spectrum}, so this scenario is not favored by our lattice results. 
\end{itemize}
In summary, Figure \ref{fig:pm_nuk_energy_spectrum} shows that no additional eigen-state or significant energy shift is observed with respect to non-interacting $N$ and $\eta_c$.  We  conclude that there is no strong indication for a $P_c$ resonance in one-channel approximation for $N\eta_{c}$ scattering.

 \subsection{Comparison with previous lattice simulations}
           
           Finally, we compare our conclusions with previous lattice simulations of $NJ/\psi$ and $N\eta_c$ systems.
           
            The s-wave interaction between a nucleon and $J/\psi(\eta_c)$ was studied using the   L{\"u}scher formalism in dynamical \cite{Liu:2008rza_0810.5412} and quenched  \cite{ hep-lat_0605009_Yokokawa:2006td} QCD.  All calculated scattering lengths $a_0$, are consistent with zero within $1$ or $2$ sigma, implying very small attractive interaction.   As $a_0 \propto \Delta E$, these results  agree with our result $\Delta E \approx 0$ within errors.

           The NPLQCD collaboration observed the  energy shift  $\Delta E\approx -20~ \text{MeV}$   of $N(0)\eta_c(0)$  ground state for  very heavy $m_\pi\simeq 800$ MeV in  $J=\frac{1}{2}^-$ ($G_1^-$) channel, which was almost independent of the volume $L\simeq 3.4 - 6.7$ fm  \cite{Netac}.  We can neither confirm neither  refute such an energy-shift given the errors of our present calculation and different $m_\pi$. This study concludes that $N\eta_c$ bound state $20~$MeV below threshold exists at such a heavy pion mass. Further accurate studies are 
          needed to explore possible existence of such a bound state at physical quark masses. If this state is theoretically confirmed, looking for its experimental signatures will be of prime interest. 
          
   Two studies 
 considered potential between $N$ and $M$ ($M=J/\psi$ or $\eta_c$) in s-wave as a function of distance withing the HALQCD method 
 in  a dynamical   \cite{Jpsi_nucleon_HALQCD_method} and a quenched  \cite{Kawanai:2010ev,Kawanai:2010cq_1007.1515}  simulation. The light-quark mass was  larger than physical with the nucleon mass $m_N\simeq 1.8~$GeV in \cite{Jpsi_nucleon_HALQCD_method} and  $m_\pi=640-870~$MeV in \cite{Kawanai:2010ev}.  
They find weakly-attractive interaction near threshold in three channels explored: $J^P=\frac{1}{2}^-,\frac{3}{2}^-$ for $NJ/\psi$ and $\frac{1}{2}^-$ for $N\eta_c$.   The resulting interaction was not  strong enough to form bound states or resonances, but the most interesting experimental region $4.3-4.5~$GeV was not explored.  The absence of a very pronounced interactions in this system agrees with our conclusions. 

Finally, the study \cite{1608.06537_Alberti2016} of hadroquarkonium picture considered the 
  static $\bar cc$  ($m_c\to \infty$) as function of distance between $\bar c$ and $c$  in presence of the nucleon. The $N_f=2+1$ CLS ensemble at $m_\pi\simeq 223~$MeV was employed.  
  The shift of the potential due to the presence of the nucleon was extracted by an impressive precision. This shifts was found to be down  only by  a few MeV.  Such a shift is compatible with our results given the uncertainties on our eigen-energies.  
  
Lattice results \cite{hep-lat_0605009_Yokokawa:2006td,Liu:2008rza_0810.5412, Kawanai:2010ev, Kawanai:2010cq_1007.1515}  were used for  determination of parameters in effective field theory \cite{Castella2018_1803.05412}. No signs of  quarkonium-nucleon bound state is found within range of applicability for described EFT.

\section{Conclusions and Outlook} \label{sec:8} 

We perform a $N_f=2$  lattice QCD simulation of $NJ/\psi$ and $N\eta_{c}$ scattering in the one-channel approximation, where $N$ denotes a proton or a neutron.  
This is the first study that reaches the energies, where the charmed pentaquarks $P_c$ resonances were observed in $NJ/\psi$ decay by the  LHCb experiment.  The resulting energies of eigenstates  in Figures  \ref{fig:vm_nuk_energy_spectrum}    and \ref{fig:pm_nuk_energy_spectrum}  are compared to the analytic predictions of (i) a scenario with non-interacting nucleon-charmonium  system and (ii) a scenario featuring a $P_c$ resonance coupled to a single channel. The non-interacting spectrum $E=E_N(p)+E_{\bar cc}(-p)$ with $\mathbf{p}=\mathbf{n}2\pi/{\tt L}$ is discrete due to periodic  boundary conditions on the lattice of  finite size ${\tt L}$.  We find that the extracted lattice spectra is consistent with the prediction of an almost non-interacting nucleon-charmonium system within errors of our calculation. The scenario based on a Breit-Wigner-type $P_c$ resonance, coupled solely to $NJ/\psi$ or to $N\eta_c$, is  not supported by our lattice data. 
The results indicate that the  existence of $P_c$ resonance  within a one-channel   
          scattering  is not favored in QCD. This might  suggest that the strong coupling between the $NJ/\psi$ with other two-hadron channels might be responsible for the existence of the $P_c$ resonances in experiment.  Future lattice simulations of coupled-channel scattering   are needed to investigate this hypothesis. 

One of the challenges in extracting the eigenstates of $NJ/\psi$ system in the current simulation is related to the high number of almost-degenerate eigenstates. This degeneracy arises due to the non-zero spins of $N$ and $J/\psi$, since a number of different partial waves $(\ell,S)$ can couple to a certain channel $J^P$. Furthermore, several $J^P$ contribute  to a certain lattice irreducible representation due to the reduced symmetry on the lattice. As a result, 
the non-interacting scenario predicts up to six degenerate linearly-independent $N(p)J/\psi(-p)$  eigenstates with the relative momentum $p^2=2(\tfrac{2\pi}{\tt L})^2$ for a given row and a given lattice irreducible representation. We establish all    such  eigenstates with $p^2\leq 2(\tfrac{2\pi}{\tt L})^2$, together with the pattern of degeneracies expected from the non-interacting case. The use of carefully-constructed interpolators  \cite{operators_2017,Callat_partial_wave} were vital for this. 

Our work is only the first step towards exploring  the dynamics of the charmed pentaquark channels  by means of the lattice QCD simulations.  Although $P_c\simeq uud\bar cc$ resonances were experimentally  observed so far only in the proton-$J/\psi$ decay, they are allowed to strongly  decay to a nucleon and a charmonium as well as to a charmed meson and a charmed baryon. The coupled-channel lattice study of  the relevant channels is a challenging task left for the future simulations. These will contain also the Wick contractions connecting a meson and a baryon, which are absent for the case of a nucleon-charmonium system considered here. The number of eigenstates in a given irreducible representation will become even denser than in the present study. The extraction of the small non-zero energy shifts will require a particular effort in reducing the statistical error, especially for the underlying baryonic component at zero and non-zero relative momenta.  The study of the systems with the non-zero total momentum will render additional information on the scattering matrices, but poses additional challenges since spin $J$ and parity $P$ are  no longer good quantum numbers, even in the continuum. Therefore, understating the realistic $P_c$ resonances through the energies of eigenstates by means of a rigorous L\"uscher-type approach  brings considerable challenges.  In light of this, it would be valuable to explore if there are any other ways to investigate these interesting systems by means of the first-principle lattice QCD. \\

\textbf{Acknowledgments} 

\vspace{0.2cm}

We are grateful to M. Padmanath for valuable   discussions and help concerning the cross-checks for the nucleon. We would like to kindly thank D. Mohler and C.B. Lang for allowing us to use the quark perambulators, generated during our previous joint projects. We acknowledge A. Hasenfratz for sharing with us the gauge configurations employed here.  We  thank   L. Leskovec for sharing GEVP code for an independent cross-check.     This work was supported by Research Agency ARRS (research core funding No. P1-0035 and No. J1-8137) and DFG grant No. SFB/TRR 55. 

\vspace{0.3cm}
{\it Note added. }- Recently, LHCb reported a discovery of a new pentaquark state  $P_c^+(4312)$ that is also observed in the $pJ/\psi $ invariant mass  \cite{LHCb_Pc_2019}.
    All the discovered  narrow  $P_c$   lie near  $\Sigma_c^+ \bar D^{0(*)}$ threshold,  indicating that coupling of $pJ/\psi $ to this channel   might be important for giving rise to $P_c$ in experiment.  This   might provide a  possible explanation as to why $P_c$ resonances are not observed in our lattice study of $pJ/\psi  $ channel in the approximation where it is decoupled from all other channels. Recent LHCb results are in line with our conclusion that the coupling of $pJ/\psi $ with other channels  might be crucial for  the existence of $P_c$ resonances. 

\bibliographystyle{h-physrev4}

\appendix

\section{Possible strong decay channels for $P_{c}^{+}$}
\label{ch:possible_decays}

 Possible two-hadron strong decay channels for $P_{c}$ are listed in Table \ref{tab:particles_pentaquark}, together with threshold locations in experiment.  Spin and parities of mesons and baryons are listed as well as the corresponding value of angular momentum $\ell$ to obtain $P_{c}$  in a given channel with quantum number $J^P$.  In Table \ref{tab:particles_pentaquark}  all possible channels for angular momentum $\ell \leq 2$ are listed. 
\begin{table*}
	\centering
	\begin{tabular}{c c c c c c c}
		\hline
		\hline
		$J^{P}$ & $\ell$ & $m_{m}+m_{b} $ & \text{meson} & $J^{PC}_{\text{meson}} $  & \text{baryon} & $J^{P}_{\text{baryon}} $\\
		&  & \text{ [MeV]} &  &  &  &  \\
		\hline
		\hline
		$\frac{3}{2}^{-}$  & $0$ & $4034$ & $J/\psi $&  $1^{-+} $  & $p$ & $\frac{1}{2}^{+} $\\ 

									&  			   & $4464$  & $D^{-*} $						  &  $1^{-+} $  & $\Sigma_c^{++}(2455)$ & $\frac{1}{2}^{+} $\\ 
									&  			   & $4387$ &  $D^{-} $						  &  $0^{-+} $  & $\Sigma_c^{++}(2520)$ & $\frac{3}{2}^{+} $\\ 
									&  			   & $4528$ & $D^{-*} $						  &  $1^{-+} $  & $\Sigma_c^{++}(2520)$ & $\frac{3}{2}^{+} $\\

									& $1$  & $4352$ & $\chi_{c0} $&  $0^{++} $  & $p$ & $\frac{1}{2}^{+} $\\ 
									& 			   & $4448$ & $\chi_{c1} $&  $1^{++} $  & $p$ & $\frac{1}{2}^{+} $\\ 						
							& $2$  & $3921$ & $\eta_c $&  $0^{-+} $  & $p$ & $\frac{1}{2}^{+} $\\ 
							&				& $4034$ & $J/\psi $&  $1^{-+} $  & $p$ & $\frac{1}{2}^{+} $\\ 

							&  			   & $4323$ &  $D^{-} $						  &  $0^{-+} $  & $\Sigma_c^{++}(2455)$ & $\frac{1}{2}^{+} $\\ 
							&  			   & $4464$  & $D^{-*} $						  &  $1^{-+} $  & $\Sigma_c^{++}(2455)$ & $\frac{1}{2}^{+} $\\ 
							&  			   & $4387$ &  $D^{-} $						  &  $0^{-+} $  & $\Sigma_c^{++}(2520)$ & $\frac{3}{2}^{+} $\\ 
							&  			   & $4528$ & $D^{-*} $						  &  $1^{-+} $  & $\Sigma_c^{++}(2520)$ & $\frac{3}{2}^{+} $

			\end{tabular}
		\begin{tabular}{c c c c c c c}
			\hline
			\hline
			$J^{P}$ & $\ell$ & $m_{m}+m_{b} $ & \text{meson} & $J^{PC}_{\text{meson}} $  & \text{baryon} & $J^{P}_{\text{baryon}} $\\
			&  & \text{ [MeV]} &  &  &  &  \\
			\hline
		\hline
		$\frac{3}{2}^{+}$ & $0$ & $4448$ & $\chi_{c1} $&  $1^{++} $  & $p$ & $\frac{1}{2}^{+} $\\ 
				
									& $1$  & $3921$ & $\eta_c $&  $0^{-+} $  & $p$ & $\frac{1}{2}^{+} $\\ 
									&				& $4034$ & $J/\psi $&  $1^{-+} $  & $p$ & $\frac{1}{2}^{+} $\\ 
									&  			   & $4323$ &  $D^{-} $&  $0^{-+} $  & $\Sigma_c^{++}(2455)$ & $\frac{1}{2}^{+} $\\ 
									&  			   & $4464$  & $D^{-*} $ &  $1^{-+} $  & $\Sigma_c^{++}(2455)$ & $\frac{1}{2}^{+} $\\ 
									&  			   & $4387$ &  $D^{-} $	 &  $0^{-+} $  & $\Sigma_c^{++}(2520)$ & $\frac{3}{2}^{+} $\\ 
									&  			   & $4528$ & $D^{-*} $& $1^{-+} $  & $\Sigma_c^{++}(2520)$ & $\frac{3}{2}^{+} $\\ 
	& $2$  & $4352$ & $\chi_{c0} $&  $0^{++} $  & $p$ & $\frac{1}{2}^{+} $\\ 
& 			   & $4448$ & $\chi_{c1} $&  $1^{++} $  & $p$ & $\frac{1}{2}^{+} $ 
		
		\end{tabular}			
\begin{tabular}{c c c c c c c}
	\hline
	\hline
$J^{P}$ & $\ell$ & $m_{m}+m_{b} $ & \text{meson} & $J^{PC}_{\text{meson}} $  & \text{baryon} & $J^{P}_{\text{baryon}} $\\
&  & \text{ [MeV]} &  &  &  &  \\
\hline
\hline
		$\frac{5}{2}^{+}$ 	&$1$ & $4034$ & $J/\psi $&  $1^{-+} $  & $p$ & $\frac{1}{2}^{+} $\\ 
									&  			   & $4464$  & $D^{-*} $&  $1^{-+} $  & $\Sigma_c^{++}(2455)$ & $\frac{1}{2}^{+} $\\ 
									&  			   & $4387$ &  $D^{-} $  &  $0^{-+} $  & $\Sigma_c^{++}(2520)$ & $\frac{3}{2}^{+} $\\ 
									&  			   & $4528$ & $D^{-*} $  &  $1^{-+} $  & $\Sigma_c^{++}(2520)$ & $\frac{3}{2}^{+} $\\ 
										& $2$  & $4352$ & $\chi_{c0} $&  $0^{++} $  & $p$ & $\frac{1}{2}^{+} $\\ 
									& 			   & $4448$ & $\chi_{c1} $&  $1^{++} $  & $p$ & $\frac{1}{2}^{+} $ 
	\end{tabular}
\begin{tabular}{c c c c c c c}
	\hline
	\hline
	$J^{P}$ & $\ell$ & $m_{m}+m_{b} $ & \text{meson} & $J^{PC}_{\text{meson}} $  & \text{baryon} & $J^{P}_{\text{baryon}} $\\
	&  & \text{ [MeV]} &  &  &  &  \\
	\hline
	\hline
	$\frac{5}{2}^{-}$ &  $0$	  	   & $4528$ & $D^{-*} $						  &  $1^{-+} $  & $\Sigma_c^{++}(2520)$ & $\frac{3}{2}^{+} $\\ 
	& 		$1$ 	   & $4448$ & $\chi_{c1} $&  $1^{++} $  & $p$ & $\frac{1}{2}^{+} $\\ 
	
	& $2$  & $3921$ & $\eta_c $&  $0^{-+} $  & $p$ & $\frac{1}{2}^{+} $\\ 
	&				& $4034$ & $J/\psi $&  $1^{-+} $  & $p$ & $\frac{1}{2}^{+} $\\ 
	&  			   & $4323$ &  $D^{-} $ &  $0^{-+} $  & $\Sigma_c^{++}(2455)$ & $\frac{1}{2}^{+} $\\ 
	&  			   & $4464$  & $D^{-*} $&  $1^{-+} $  & $\Sigma_c^{++}(2455)$ & $\frac{1}{2}^{+} $\\ 
	&  			   & $4387$ &  $D^{-} $ &  $0^{-+} $  & $\Sigma_c^{++}(2520)$ & $\frac{3}{2}^{+} $\\ 
	&  			   & $4528$ & $D^{-*} $	 &  $1^{-+} $  & $\Sigma_c^{++}(2520)$ & $\frac{3}{2}^{+} $
\end{tabular}
	\caption{Possible decay channels of $P_{c}$ for various choices of $J^{P}$  and angular momentum $\ell\leq 2$ .  }  
	\label{tab:particles_pentaquark}
\end{table*}

\section{List of quantum numbers for linearly independent operators}
\label{ch:appA}

The lists of employed linearly independent $NJ/\psi$ and $N\eta_c$ operators $O^{[J,\ell,S]}_\Gamma$ are given for all irreps $\Gamma$.  Here  in Table \ref{tab:qn_pm_nuk} and Table \ref{tab:qn_vm_nuk}  $(J,\ell,S)$ indicate continuum quantum numbers  of the operators before subduction to $\Gamma$. The number of linearly independent operators is equal to the number of eigen-states in the non-interacting limit, as discussed in the main text. 

\begin{table*}
		\centering
	\begin{tabular}{c | c c c c}
	irrep   & $p^2$ & $J$ & $\ell$ & $S$ \\
	\hline
	\hline
	$G_1^-$  & $0$ & $\frac{1}{2}$ & $0$ & $\frac{1}{2}$ \\ 
				   & $1$ & $\frac{1}{2}$ & $0$ & $\frac{1}{2}$ \\ 
				   & $2$ & $\frac{1}{2}$ & $0$ & $\frac{1}{2}$  \\
	\hline
	$G_1^+$  & $1$ & $\frac{1}{2}$ & $1$ & $\frac{1}{2}$ \\ 
				  & $2$ & $\frac{1}{2}$ & $1$ & $\frac{1}{2}$ \\ 
		\hline
	$G_2^-$  & $2$ & $\frac{5}{2}$ & $2$ & $\frac{1}{2}$ \\ 
		\hline
	$G_2^+$  & $2$ & $\frac{5}{2}$ & $3$ & $\frac{1}{2}$ 
		  			\end{tabular} 
\begin{tabular}{c  | c c c c}
irrep  & $p^2$ & $J$ & $\ell$ & $S$ \\
\hline 
\hline	
	$H^-$ & $1$ & $\frac{3}{2}$ & $2$ & $\frac{1}{2}$\\
				 & $ 2$ & $\frac{3}{2}$ & $2$ & $\frac{1}{2}$ \\ 
			 & $2$ & $\frac{5}{2}$ & $2$ & $\frac{1}{2}$  \\
		\hline
	$H^+$	 & $ 1$ & $\frac{3}{2}$ & $1$ & $\frac{1}{2}$ \\ 
					 & $2$ & $\frac{3}{2}$ & $1$ & $\frac{1}{2}$\\ 
				 & $2$ & $\frac{3}{2}$ & $3$ & $\frac{1}{2}$  
	\end{tabular}
\caption{Combinations of quantum numbers for linearly independent operators after subduction to chosen irrep for $N\eta_{c}$ scattering. }
\label{tab:qn_pm_nuk}
\end{table*}

\begin{table*}
		\centering
	\begin{tabular}{c | c c c c}
		irrep  & $p^2$ & $J$ & $\ell$ & $S$ \\
		\hline
		\hline
		$G_1^-$  & $0$ & $\frac{1}{2}$ & $0$ & $\frac{1}{2}$ \\ 
	 & $1$ & $\frac{1}{2}$ & $0$ & $\frac{1}{2}$ \\ 
	 & $1$ & $\frac{1}{2}$ & $2$ & $\frac{3}{2}$  \\
		 & $2$ & $\frac{1}{2}$ & $0$ & $\frac{1}{2}$ \\ 
	 & $2$ & $\frac{1}{2}$ & $2$ & $\frac{3}{2}$ \\ 
		 & $2$ & $\frac{7}{2}$ & $2$ & $\frac{3}{2}$  \\
		\hline
		$G_1^+$  & $1$ & $\frac{1}{2}$ & $1$ & $\frac{1}{2}$ \\ 
					  & $1$ & $\frac{1}{2}$ & $1$ & $\frac{3}{2}$ \\ 
					   & $2$ & $\frac{1}{2}$ & $1$ & $\frac{1}{2}$  \\
					  & $2$ & $\frac{1}{2}$ & $1$ & $\frac{3}{2}$ \\ 
					 & $2$ & $\frac{7}{2}$ & $3$ & $\frac{3}{2}$ 
		  			\end{tabular} 
\begin{tabular}{c  | c c c c}
irrep  & $p^2$ & $J$ & $\ell$ & $S$ \\
\hline 
\hline	
		$G_{2}^{-}$  & $1$ & $\frac{5}{2}$ & $2$ & $\frac{3}{2}$   \\
						    & $2$ & $\frac{5}{2}$ & $2$ & $\frac{1}{2}$  \\
	 					   & $2$ & $\frac{5}{2}$ & $2$ & $\frac{3}{2}$  \\
	   					    & $2$ & $\frac{5}{2}$ & $4$ & $\frac{3}{2}$ \\ 
	    \hline
	   	$G_{2}^{+}$  & $1$ & $\frac{5}{2}$ & $1$ & $\frac{3}{2}$   \\
	   					 & $2$ & $\frac{5}{2}$ & $3$ & $\frac{1}{2}$  \\
	   					& $2$ & $\frac{5}{2}$ & $1$ & $\frac{3}{2}$  \\
	   						& $2$ & $\frac{5}{2}$ & $3$ & $\frac{3}{2}$ 
	  			\end{tabular} 
	\begin{tabular}{c  | c c c c}
		irrep  & $p^2$ & $J$ & $\ell$ & $S$ \\
  \hline 
  \hline	
	   $H^{-}$ &  $0$ & $\frac{3}{2}$ & $0$ & $\frac{3}{2}$   \\
				 & $1$ &  $\frac{3}{2}$ & $2$ & $\frac{1}{2}$ \\
				 & $1$ & $ \frac{3}{2}$ & $ 0$ & $ \frac{3}{2}$   \\
				 & $1$ & $ \frac{3}{2}$ & $ 2$ & $ \frac{3}{2}$  \\ 
				 & $2$ & $ \frac{3}{2}$ & $ 0$ & $ \frac{3}{2}$   \\
				 & $2$ & $\frac{3}{2}$ & $2$ & $\frac{3}{2}$   \\
				 & $2$ & $\frac{5}{2}$ & $2$ & $\frac{1}{2}$   \\
				 & $2$ & $\frac{5}{2}$ & $2$ & $\frac{3}{2}$   \\
				 & $2$ & $\frac{5}{2}$ & $4$ & $\frac{3}{2}$  \\
				 & $2$ & $\frac{7}{2}$ & $2$ & $\frac{3}{2}$  
		  			\end{tabular} 
\begin{tabular}{c  | c c c c}
irrep  & $p^2$ & $J$ & $\ell$ & $S$ \\
\hline 
\hline	
  		$H^{+}$  & $1$ & $\frac{3}{2}$ & $1$ & $\frac{1}{2}$ \\
	 			       & $1$ & $\frac{3}{2}$ & $1$ & $\frac{3}{2}$  \\
					  & $1$ & $\frac{3}{2}$ & $3$ & $\frac{3}{2}$ \\
	 	      		 & $2$ & $\frac{3}{2}$ & $1$ & $\frac{1}{2}$  \\ 
					 & $2$ & $\frac{3}{2}$ & $3$ & $\frac{3}{2}$  \\
				 & $2$ & $\frac{5}{2}$ & $3$ & $\frac{1}{2}$\\
				& $2$ & $ \frac{5}{2}$ & $1$ & $\frac{3}{2}$ \\
				 & $2$ & $ \frac{5}{2}$ & $3$ & $\frac{3}{2}$ \\
				 & $2$ & $ \frac{7}{2}$ & $3$ & $\frac{1}{2}$ 
	\end{tabular}
	\caption{Combinations of quantum numbers for linearly independent operator-types after subduction to chosen irrep for $NJ/\psi$ scattering. }
	\label{tab:qn_vm_nuk}
\end{table*}

\section{Resulting energies  of eigenstates for  $N\eta_c$ and $NJ/\psi$}

Resulting lattice eigen-energies in both $N\eta_c$ and $NJ/\psi$ channels are presented in Table \ref{tab:pm_nuk_fit} and \ref{tab:vm_nuk_fit}.

\begin{table*}
	\centering
\begin{tabular}{cc|ccc}

	irrep & state $n$ & $E_na$ &  $\sigma_{E_n}a$ \\ 
	\hline 
	\hline
	$G_{1}^{-}$ & $1$  & $2.183$ &$ 0.022$  \\ 
 
	& $2$ &  $2.286$ & $0.023$ \\ 

	& $3$ &  $2.455$ & $0.027$  \\ 
	\hline 
	$G_{1}^{+}$ & $1$  & $2.291$ & $0.020$ \\ 
 
	& $2$ & $2.446$ & $0.026$ \\ 
	\hline 
	$G_{2}^{-}$ & $1$ & $2.448$ & $0.028$ \\ 
	\hline 
	$G_{2}^{+}$ & $1$  & $2.452$ & $0.026$
			\end{tabular} 
\begin{tabular}{cc|cc}
irrep & state $n$ &  $E_na$ & $\sigma_{E_n}a$ \\ 
\hline 
\hline
	$H^{-}$ & $1$  & $2.294$ & $0.020$ \\ 

	& $2$  & $2.438$ & $0.015$  \\ 

	& $3$ & $2.456$ & $0.030$  \\ 
	\hline 
	$H^{+}$ & $1$ & $2.294$ & $0.020$  \\ 

	& $2$ &$2.450$ & $0.029$ \\ 

	& $3$ & $2.447$ & $0.024$ 
\end{tabular} 
\caption{Eigen-energies $E_n$ and their errors $\sigma_{E_n}$ for  the $N\eta_c$ system extracted  from one-exponential correlated fits in $t=[7,10]$. }
\label{tab:pm_nuk_fit}
\end{table*}

\begin{table*}
\centering
	\begin{tabular}{cc|cc}
		irrep & state $n$ &  $E_na$ & $\sigma_{E_n}a$ \\ 
		\hline 
		\hline
		$G_{1}^{-}$ & $ 1$ & $ 2.253$ & $0.022$ \\ 
		& $ 2$ &$ 2.351$ & $0.020$ \\ 
		& $ 3$ &  $ 2.360$ & $0.021$  \\ 
		& $ 4$ & $ 2.511$ & $0.028$  \\ 
		& $ 5$ &  $ 2.510$ & $0.027$ \\ 
		& $ 6$ &  $ 2.518$ & $0.029$ \\ 
		\hline
		$G_{1}^{+}$ & $ 1$ &$ 2.358$ & $0.020$ \\ 
		& $ 2$ &  $ 2.354$ & $0.021$   \\ 
		& $ 3$ & $ 2.503$ & $0.027$ \\ 
		& $ 4$ & $ 2.511$ & $0.027$  \\ 
		& $ 5$ & $ 2.507$ & $0.030$   
			\end{tabular} 
\begin{tabular}{cc|cc}
irrep & state $n$ &  $E_na$ & $\sigma_{E_n}a$ \\ 
\hline 
\hline
		$G_{2}^{-}$& $ 1$  & $ 2.357$ & $0.021$ \\ 
		& $ 2$ & $ 2.492$ & $0.030$ \\ 
		& $ 3$ & $ 2.514$ & $0.028$ \\ 
		& $ 4$  & $ 2.506$ & $0.028$ \\ 
		\hline 
		$G_{2}^{+}$& $ 1$ & $ 2.357$ & $0.020$ \\ 
		& $ 2$ & $ 2.496$ & $0.035$ \\ 
		& $ 3$ & $ 2.523$ & $0.026$  \\ 
		& $ 4$ & $ 2.516$ & $0.034$  
			\end{tabular} 
		\begin{tabular}{cc|cc}
	irrep & state $n$ &  $E_na$ & $\sigma_{E_n}a$ \\ 
			\hline 
			\hline
		$H^{-}$& 
		$ 1$ &  $ 2.259$ & $0.023$  \\ 
		& $ 2$ & $ 2.362$ & $0.015$  \\ 
		& $ 3$ &  $ 2.371$ & $0.016$ \\ 
		& $ 4$ & $ 2.374$ & $0.016$  \\ 
		& $ 5$ &  $ 2.517$ & $0.025$  \\ 
		& $ 6$ &  $ 2.520$ & $0.030$ \\ 
		& $ 7$ &  $ 2.519$ & $0.023$ \\ 
		& $ 8$ &$ 2.530$ & $0.024$ \\ 
		& $ 9$ & $ 2.527$ & $0.023$  \\ 
		& $10$ & $2.532$ & $ 0.023$   
			\end{tabular} 
\begin{tabular}{cc|cc}
irrep & state $n$ &  $E_na$ & $\sigma_{E_n}a$ \\ 
\hline 
\hline
		$H^{+}$ 
		& $ 1$ & $ 2.369$ & $0.015$ \\ 
		& $ 2$ &  $ 2.371$ & $0.015$ \\ 
		& $ 3$ &  $ 2.375$ & $0.016$ \\ 
		& $ 4$ & $ 2.523$ & $0.020$ \\ 
		& $ 5$ & $ 2.524$ & $0.020$ \\ 
		& $ 6$ & $ 2.516$ & $0.027$  \\ 
		& $ 7$ & $ 2.527$ & $0.021$ \\ 
		& $ 8$ & $ 2.520$ & $0.034$  \\ 
		& $ 9$ & $ 2.529$ & $0.022$ 
	\end{tabular} 
\caption{Eigen-energies $E_n$ and their errors $\sigma_{E_n}$ for  the $NJ/\psi$ system extracted  from one-exponential correlated fits in $t=[7,10]$.    }
	\label{tab:vm_nuk_fit}
\end{table*}

\end{document}